\documentclass[a4paper,fleqn]{cas-sc}

\usepackage[authoryear]{natbib}
\usepackage[section]{placeins} %PACKAGE I ADDEDTO FIX IMAGE POSITIONS!
\usepackage{booktabs,adjustbox,longtable} %For the annex table
\usepackage{lineno}
\def\tsc#1{\csdef{#1}{\textsc{\lowercase{#1}}\xspace}}
\tsc{WGM}
\tsc{QE}
\begin{document}
\let\WriteBookmarks\relax
\def\floatpagepagefraction{1}
\def\textpagefraction{.001}

\shorttitle{}    

\shortauthors{Duarte Branco et al.}  

\title[mode = title]{Dynamical origin of Theia, the last giant impactor on Earth}  




\author[1,2]{Duarte Branco}[orcid=0009-0001-8348-1318]%[<options>]


%\ead{}


\credit{Writing - Original Draft, Writing - Review \& Editing, Investigation, Methodology, Software, Formal analysis, Visualization}

\affiliation[1]{organization={Faculdade de Ciências da Universidade de Lisboa},
            addressline={Campo Grande 016}, 
            city={Lisbon},
%          citysep={}, % Uncomment if no comma needed between city and postcode
            postcode={1749-016}, 
            country={Portugal}}

\affiliation[2]{organization={Institute of Astrophysics and Space Sciences},
            addressline={Observatório Astronómico de Lisboa, Tapada da Ajuda, Edifício Leste - 2º, Piso}, 
            city={Lisbon},
%          citysep={}, % Uncomment if no comma needed between city and postcode
            postcode={1349-018}, 
            country={Lisbon}}

\author[1,2]{Pedro Machado}[orcid=0000-0003-1760-3871]%[]

% Footnote of the second author

% Email id of the second author
\ead{pmmachado@fc.ul.pt}


\credit{Funding acquisition, Supervision, Project administration, Resources}


\cormark[1]
\cortext[1]{Corresponding author at: Faculdade de Ciências da Universidade de Lisboa, Campo Grande 016, 1349-018 Lisbon, Portugal}

%\fntext[2]{}

\author[3]{Sean N. Raymond}[orcid=0000-0001-8974-0758]%[]

% Footnote of the second author
%\fnmark[3]

% Email id of the second author
%\ead{sean.raymond@u-bordeaux.fr}


\credit{Supervision, Project administration, Writing - Review \& Editing, Conceptualization, Methodology}

\affiliation[3]{organization={Laboratoire d'Astrophysique de Bordeaux, Univ. Bordeaux, CNRS},
            addressline={All. Geoffroy Saint-Hilaire}, 
            city={Bordeaux},
%          citysep={}, % Uncomment if no comma needed between city and postcode
            postcode={33600 Pessac}, 
            country={France}}


\newcommand{\mearth}{M_{\oplus}} %adding a new command in since you write \mearth a few times
%\linenumbers
\begin{abstract} %Full Check
Cosmochemical studies have proposed that Earth accreted roughly 5-10\% of its mass from carbonaceous (CC) material, with a large fraction delivered late via its final impactor, Theia (the Moon-forming impactor). Here, we evaluate this idea using dynamical simulations of terrestrial planet formation, starting from a standard setup with a population of planetary embryos and planetesimals laid out in a ring centered between Venus and Earth's orbits, and also including a population of CC planetesimals and planetary embryos scattered inward by Jupiter.  We find that this scenario can match a large number of constraints, including i) the terrestrial planets' masses and orbits; ii) the CC mass fraction of Earth; iii) the much lower CC mass fraction of Mars, as long as Mars only accreted CC planetesimals (but no CC embryos); iv) the timing of the last giant (Moon-forming) impact; and v) a late accretion phase dominated by non-carbonaceous (NC) bodies.  For this scenario to work, the total mass in scattered CC objects must have been $\sim 0.2-0.3 \mearth$, with an embryo-to-planetesimal mass ratio of at least 8, and CC embryos in the $\sim 0.01-0.05 \mearth$ mass range. In that case, our simulations show there are roughly 50-50 odds of Earth's last giant impactor (Theia) having been a carbonaceous object -- either a pure CC embryo or an NC embryo that previously accreted a CC embryo. Our simulations thus provide dynamical validation of cosmochemical studies. % Ratio changed to "at least 8" due to the results being closer to 8 than 10
    
%    Cosmochemical studies have proposed that Earth accreted roughly 5-10\% of its mass from carbonaceous material, with a large fraction delivered late via its final impactor, Theia (the Moon-forming impactor). 
%    Here, we evaluate this idea using dynamical simulations of terrestrial planet formation, starting from a standard setup with a population of planetary embryos and planetesimals laid out in a ring centered between Venus and Earth's orbits, and also including a population of carbonaceous planetesimals and planetary embryos scattered inward by Jupiter. 
%    We find that Earth and Venus generally accrete significantly more carbonaceous material than Mars, while having dryer late accretions. 
%    We find scenarios that are consistent with the cosmochemical predictions -- in roughly half of simulations, Theia contained a considerable CC component. %and in another 23\% it was an agglomeration that included at least one carbonaceous embryo. 
%    The size of the CC component and how often this ocurred in simulations depends on the mass distribution of carbonaceous material crossing the terrestrial planet-forming region. \nocite{*}
    \end{abstract}


\begin{highlights}
\item We dynamically test the idea, derived from cosmochemical studies, that Theia may have been a carbonaceous (CC) object~\citep{Budde2019,Nimmo2024}
\item We ran N-body simulations of the late stages of terrestrial planet growth, including a tail of carbonaceous objects assumed to have been scattered inward during Jupiter and Saturn's accretion.
\item Our dynamical simulations show that this scenario is plausible, as in roughly 50\% of viable systems, Earth's last giant impactor is either a pure carbonaceous planetary embryo or a non-carbonaceous embryo that previously accreted carbonaceous material.
\item We constrain the system parameters (CC mass and embryo:planetesimal mass ratio) for which a carbonaceous last giant impactor is viable.
\end{highlights}

% Keywords

\maketitle

% Main text
\section{Introduction}\label{cha:introduction}%%%%%%%%%%%%%%%%%%%%%%%%%%%%%%%%%%%%%%%%%%%%%%%%%%%%%%%%%%%%%%%%%%%%%%%%%%%%%%%%%%%%%%%% INTRODUCTION

The standard scenario of terrestrial planet formation invokes several phases of growth~\citep[for a review, see][]{raymond22}.  First, planetesimals form from concentrations of pebbles or large dust particles that either drifted inward or condensed in preferential locations~\citep[e.g.][]{johansen14,birnstiel16,drazkowska16}.  Current modeling favors condensation fronts as preferred locations for planetesimal formation, and infer that the Solar System likely formed an inner ring of planetesimals at $\sim 1$~au~\citep{drazkowska18,lichtenberg21,Morbidelli2022,izidoro22}.  Next, Moon-to-Mars-sized planetary embryos grow by accreting planetesimals~\citep[e.g.][]{KokuboIda1998,kokubo00,leinhardt05,wetherill93,weidenschilling97}. After, there was a chaotic phase of giant impacts between planetary embryos, in the presence of a population of remaining planetesimals~\citep[e.g.][]{Wetherill1990,CHAMBERS1998304,Raymond2006,OBRIEN2006,Morbidelli2012,Raymond2014}.  The last giant impact on the proto-Earth was with an embryo commonly called `Theia' -- this impact caused the final differentiation event on the Earth and led to the formation of the Moon~\citep{Benz1986,Agnor1999,CanupAsphaug2001,Canup2012,Cuk2012,canup2021origin}.  Most studies invoke a timing of 30-100 Myr after CAIs for the Moon-forming impact~\citep{Kleine09,Avice2014,Jacobson2014,Thiemens2019}, although in principle it may have taken place as early as 10 Myr after CAIs~\citep{Fischer2018}. Finally, during the `late accretion' (or `late veneer' as is more commonly used in cosmochemistry) phase, Earth and the other rocky planets underwent collisions with planetesimals, which delivered highly-siderophile elements to Earth's mantle and crust~\citep{day07,walker09,bottke10,raymond13,zhu21}.  

Meteorite measurements can constrain the building blocks of the terrestrial planets~\citep[e.g.][]{dauphas17}. Given the isotopic dichotomy measured in meteorites~\citep{WARREN201193,Kruijer2020}, there was likely a corresponding dichotomy in the Solar System's planetesimal populations, with non-carbonaceous (NC) material originating interior to Jupiter's orbit and carbonaceous (CC) material exterior~\cite{KRUIJER2017}. Based on the composition of the bulk silicate Earth, it has been inferred that Earth is composed of mostly NC material, with a $\sim 4-10\%$ CC contribution~\citep[e.g.][]{steller2022,savage2022,martins23}, although the exact fraction depends on whether a CI or CV endmember is used~\citep{KLEINE2023,Nimmo2024}. In contrast, Mars appears to have had a smaller CC contribution of at most a few percent~\citep{KLEINE2023}. Measurements and modeling both suggest that the late accretion phase, which encompasses the final $\sim 0.5\%$ of material added to Earth, was dominated by NC material for both Earth and Venus~\citep{Gillmann2020,Worsham2021}. 

Carbonaceous objects are thought to have accreted beyond Jupiter's orbit~\citep{KRUIJER2017,Brasser2020} and been scattered inward during Jupiter and Saturn's gas accretion~\citep{RaymondIzidoro2017} or migration~\citep{Walsh2011,RaymondIzidoro2017,Pirani2019}. Some were trapped in the asteroid belt via aerodynamic gas drag, and other -- in particular, larger bodies that felt weaker drag -- were scattered all the way onto orbits that crossed the terrestrial planet region. Chemical modeling suggests that Earth acquired most of its carbonaceous material toward the later part of the main accretion phase~\citep{Rubie2015,Nimmo2024,Dauphas2024}. \cite{Nimmo2024} proposed that the bulk of Earth's carbonaceous material was accreted in the form of a few large impactors rather than a sea of planetesimals~\citep[a conclusion also reached by][]{JOIRET2024}. \cite{Budde2019} used Mo isotopes to propose the most extreme case, that the last giant impactor (Theia) was a carbonaceous planetary embryo. A top-heavy distribution of CC bodies in the inner Solar System has the advantage of making it easier to explain why Mars has a lower CC fraction than Earth~\citep{KLEINE2023} and also why late accretion was dominated by NC material, considering that late accretion is assumed to include only planetesimal impacts~\citep{Bottke2010,Morbidelli2015}.

In this paper, we dynamically test the idea that Theia was a carbonaceous object. We use N-body simulations to model the accretion of the terrestrial planets in conjunction with a tail of material scattered inward by Jupiter and Saturn's rapid gas accretion.  We conclude that the scenario proposed by \cite{Budde2019} and \cite{Nimmo2024} is dynamically plausible, and that Theia could have been either a CC planetary embryo or an embryo consisting of a mixture of NC and CC material.

%%%%%%%%%%%%%%%%%%%%%%%%%%%%%%%%%%%%%%%%%%%%%%%%%%%%%%%%%%%%%%%%%%%%%%%%%%%%%%%%%%%%%%%%%%%%%%%%%%%%%%%%%%%%%%%%%%%%%%%%%%%%%%%%%%%%%%%%%%%
%%%%%%%%%%%%%%%%%%%%%%%%%%%%%%%%%%%%%%%%%%%%%%%%%%%%%%%%%%%%%%%%%%%%%%%%%%%%%%%%%%%%%%%%%%%%%%%%%%%%%%%%%%%%%%%%%%%%%%%%%%%%%%%%%%%%%%%%%%%
%%%%%%%%%%%%%%%%%%%%%%%%%%%%%%%%%%%%%%%%%%%%%%%%%%%%%%%%%%%%%%%%%%%%%%%%%%%%%%%%%%%%%%%%%%%%%%%%%%%%%%%%%%%%%%%%%%%%%%%%%%%%%%%%%%%%%%%%%%%
%%%%%%%%%%%%%%%%%%%%%%%%%%%%%%%%%%%%%%%%%%%%%%%%%%%%%%%%%%%%%%%%%%%%%%%%%%%%%%%%%%%%%%%%%%%%%%%%%%%%%%%%%%%%%%%%%%%%%%%%%%%%%%%%%%%%%%%%%%%
\section{Methodology}%%%%%%%%%%%%%%%%%%%%%%%%%%%%%%%%%%%%%%%%%%%%%%%%%%%%%%%%%%%%%%%%%%%%%%%%%%%%%%%%%%%%%%%%%%%%%%%%%%%%%%%%%%%%%%%%%%%%%%%%%%%%%%%%%%%%%%%%%%%
\label{cha:Methodology}

We performed a suite of N-body simulations of the accretion of the terrestrial planets.  To start (Sec~\ref{sec:TerrOnly}), we replicate the `annulus' or `empty asteroid belt' scenario proposed by \cite{Hansen_2009} and tested by a number of subsequent studies~\citep{Kaib_2015,Raymond2017SciAdv,Nesvorny2021,Woo_2023,JOIRET2024}.  We explain the initial conditions and integration method.  Then (Sec 2.2), we include a tail of carbonaceous material assumed to have been scattered inward by Jupiter~\citep{RaymondIzidoro2017}. Finally (Sec 2.3), we explain how we addressed the effect of the giant planet instability~\citep{Tsiganis2005,Nesvorny2018} in a subset of simulations.

\subsection{Simulations to match the terrestrial planets} 
\label{sec:TerrOnly}

Our simulations start in the late phase of terrestrial accretion, after dispersion of the gaseous disk.  We make the assumption that the solid mass is divided between planetesimals and planetary embryos, and that they are initial distributed in a narrow ring, following the annulus model \citep{Hansen09,Kaib_2015,Nesvorny2021} and consistent with recent models of planetesimal formation~\citep[e.g.][]{Drazkowska2017,Morbidelli2022,Izidoro2022}.  As in previous studies~\citep[e.g.][]{CHAMBERS1998304,Raymond2009}, we assume that embryos are spaced in units of mutual Hill radii, defined as (\cite{Chambers1996}):
\begin{equation}
    R_{Hm_{i,i+1}}=\left[\frac{M_i+M_{i+1}}{3M_\odot}\right]^{1/3}\frac{(a_i+a_{i+1})}{2},
\end{equation}
\noindent where $M$ and $a$ are the mass and semi-major axis of the two planets and $M_\odot$ is the stellar mass. We assume a surface density profile that scales with orbital radius $r$ as $\Sigma(r)=\Sigma_1r^{-3/2}$~\citep[note that, for broad disks, the exponent has important consequences for accretion; see][]{Raymond_2005,Izidoro2015}. The planetesimal number and mass is the same within every set of simulations, with only their distribution changing.

We varied a few parameters to achieve a good replica of the Solar System, including the position and width of the annulus, the inter-embryo spacing (measured in units of mutual Hill radii), and the total masses of embryos and planetesimals within the annulus.  The orbital eccentricities of planetesimals and embryos were initially set to zero, and the inclination was randomly chosen between zero and 0.5$^{\circ}$, following \cite{Hansen_2009}. The other orbital angles were drawn at random. We assumed that Jupiter and Saturn were already present, and trapped in a mutual 3:2 mean motion resonance (configuration in table \ref*{tab:JS32}), consistent with simulations of their migration within the Sun's gaseous disk~\citep{Morbidelli2007,Pierens2008}.  Of course, our simulations start after dispersal of the disk so gas drag and migration are not accounted for. %changed morby 2007 to morbidelli

\begin{table}[width=.9\linewidth,cols=7,pos=h]
\caption{Initial positions for Jupiter and Saturn in simulations. Symbols represent: Semi-major axis (a), eccentricity (e), inclination (i), argument of pericentre (g), longitude of ascending node (n) and mean anomaly (M).}\label{tab:JS32}
\begin{tabular*}{\tblwidth}{@{} LLLLLLL@{} }
\toprule
Planet & a (AU) & e$\times 10^{-2}$ & i ($\deg$) & g ($\deg$) & n ($\deg$) & M ($\deg$) \\ % Table header row
\midrule
Jupiter & 5.4297 & 0.490 & 0.1 & 0 & 113.2953 & 6.9170   \\
Saturn  & 7.2987 & 0.995 & 0.2 & 10 & 276.4414 & 191.6128 \\
\bottomrule
\end{tabular*}
\end{table}

\subsubsection{Integration Method} 
Our simulations used the hybrid integrator in the \emph{Mercury6} N-Body code~\citep{Chambers1999}, which uses the mapping of \cite{Wisdom92} when objects are well-separated and switches to a Bulirsch-Stoer integrator for close encounters.  Our simulations were carried out for 200 Myr using a time step of 6 days.  The size of the Sun was inflated to 0.15 au, to avoid numerical errors during perihelion passages~\citep{Rauch_Holman99}.  The Bulirsch-Stoer accuracy parameter was set to $10^{-11}$.  Collisions were treated as inelastic mergers. 

Embryos interacted gravitationally with all other massive objects in the simulation, but planetesimal particles did not interact gravitationally with other planetesimals (but did feel the gravity of the embryos and planets).

\subsubsection{Simulation outcomes} 
To evaluate the success of simulations, we compared the orbits of the planets to the real ones', and also relied on two system-wide quantities: the terrestrial planets' radial mass concentration ($RMC$) and the normalized angular momentum deficit ($AMD$) (\cite{Chambers2001}, \cite{Lascar1997}).  These are defined as:
\begin{equation}
    AMD=\frac{\Sigma_j m_j \sqrt[]{a_j}\left[1-\sqrt[]{(1-e_j^2)}\cos i_j\right]}{\Sigma_j m_j \sqrt[]{a_j}}
    \label{eq:AMD}
\end{equation}
\begin{equation}
    RMC=\max\left(\frac{\Sigma_jm_j}{\Sigma_jm_j[\log_{10}(a/a_j)]^2}\right)
    \label{eq:RMC}
\end{equation}
\noindent  The terrestrial planets' real values are $AMD=0.0018$ and $RMC=89.9$ (\cite{Chambers2001}).

\begin{figure}[htbp]
  \centering
  \includegraphics[width=0.8\linewidth]{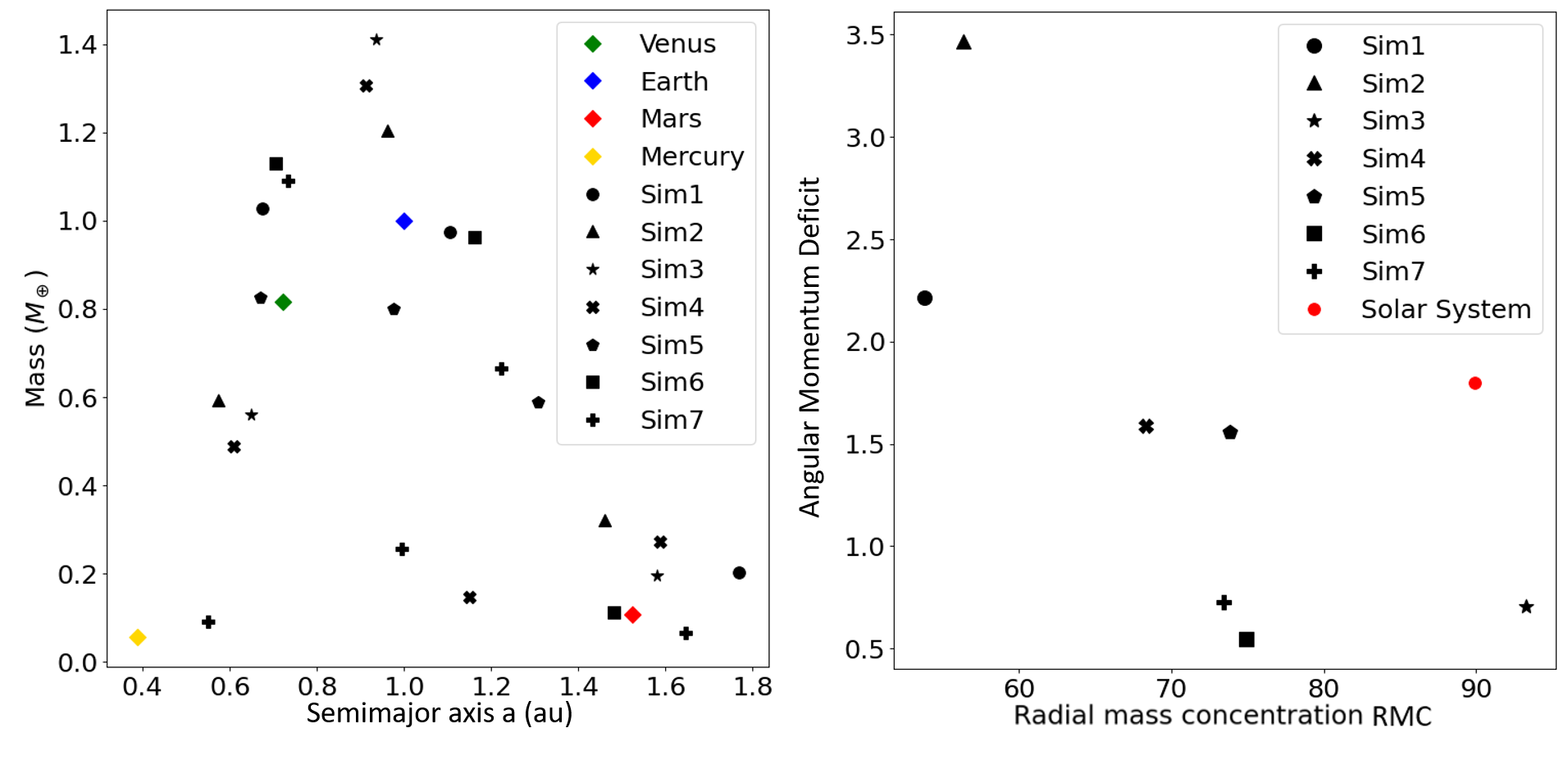}
  \caption{Comparisons of 7 simulations with no CC objects from section \ref*{sec:TerrOnly} with the real solar system. {\bf Left:} Mass and semi-major axis are compared. {\bf Right:} RMC and AMD are compared.}
  \label{fig:model2MaAMDRMC}
\end{figure}

Table \ref*{tab:model2params} lists the parameters that we found provided the best match to the terrestrial planets. Figure~\ref*{fig:model2MaAMDRMC} show the outcome of 7 simulations with this setup, in terms of the planets' orbits and the system-wide $RMC$ and $AMD$ values.

\begin{table}[width=.9\linewidth,cols=4,pos=h]
    \caption{Parameters chosen to replicate the terrestrial planets.}\label{tab:model2params}
    \begin{tabular*}{\tblwidth}{@{} LLLL@{} }
    \toprule
    Parameter  & Value & Parameter & Value \\ % Table header row
    \midrule
    Annulus radius & 0.7 to 1.2 AU & Mutual hill radii spacing ($\Delta$) & 2 to 4\\
    $\Sigma_1$ & 20.48 g cm$^{-2}$ & Number of planetesimals & 900\\
    Total embryo mass target & 2 $M_\oplus$  & Total planetesimal mass & 0.25 $M_\oplus$\\
    Embryo mass range & 0.03 $M_\oplus$ to 0.07 $M_\oplus$ & Range of embryo number & 34 to 43\\
    \bottomrule
    \end{tabular*}
\end{table}

\FloatBarrier

\subsection{Simulations including carbonaceous (CC) planetesimals} %FULL CHECK
Our next step was to introduce a population of CC objects that were scattered inwards and into intersecting orbits with the terrestrial planets during Jupiter and Saturn's rapid gas accretion. \cite{RaymondIzidoro2017} showed that gas drag plays a central role in size-sorting objects scattered inward by the gas giants.  Smaller planetesimals feel stronger gas and are trapped on wider orbits, strictly in the outer parts of the asteroid belt, whereas larger ones (typically $D \sim 1000$~km) may be scattered onto more eccentric orbits that cross within the terrestrial planet realm. If we consider an embryo/planetesimal size dichotomy, it is natural that scattered embryos would have a much higher probability of interacting with the terrestrial planets and potentially delivering CC material.  Nonetheless, given our ignorance of the mass distribution of objects scattered by Jupiter, and the fact that Jupiter's migration can also scatter planetesimals in a less size-dependent manner~\citep{Walsh2011}, we tested three different scenarios: 
\begin{itemize}
    \item Only small CC objects -- that is, planetesimals.  We call this scenario \emph{small only}.
    \item Only large CC objects -- planetary embryos. We call this scenario \emph{large only}.
    \item Both CC planetesimals and embryos. This is the \emph{mixed} scenario. 
\end{itemize}

These simulations started from the same setup described in Sec 2.1, but with an additional CC component. Carbonaceous planetesimals and embryos were distributed in a similar fashion to the setup in \cite{OBRIEN2014}, with perihelia $q$ uniformly drawn between 0.7 AU$\leq q\leq$1.5 AU, inclinations following a Rayleigh distribution with $\sigma = 2.5$, and aphelia ($Q$) less than or equal to 5.5 AU. As above, CC embryos gravitationally interact with all massive objects in the simulation, whereas CC planetesimals do not self-interact (or interact with non-carbonaceous planetesimals).

In the \emph{small only} scenario, 1050 CC planetesimals were included with a total mass of 0.2 M$_\oplus$~\citep[following][]{JOIRET2024}.  In the \emph{large only} simulations, 25 CC embryos were included with masses between 1\%-2\% of Earth's mass. The number of CC embryos is somewhat higher than one would expect if the typical CC embryo mass were the same as the often-assumed $\sim$Mars-mass, but having a significant number of particles helps in determining the odds of a collision with Earth.  

In the \emph{mixed} scenario, we made use of the collision probabilities for CC objects with the growing rocky planets from the  \emph{small only} and  \emph{large only} scenarios.  We adjusted the mass of CC objects to try to keep the mass of the terrestrial planets close to their actual values.  As such, the \emph{mixed} simulations each started with 15 CC embryos with masses between 1\%-2\% of the Earth's mass and 500 planetesimals carrying a total mass of 0.04 M$_\oplus$, for a total mass range in CCs between 0.19 M$_\oplus$-0.34 M$_\oplus$.

\subsection{The giant planet dynamical instability}
\label{sec:Methodology_GIP}
In a subset of simulations, we very simply mimicked the effect of the giant planet dynamical instability (\cite{Tsiganis2005}, \cite{Morbidelli2007}, see~\cite{Nesvorny2018} for a review). We chose 10 simulations from each of our three sets (9 from the small one) and instantaneously `jumped' the orbits of Jupiter and Saturn to their present-day ones. 
This was done at a simulation time of 20 Myr.  It is important to keep in mind that the exact timing of the instability is uncertain. A variety of cosmochemical and dynamical constraints suggest that it must have taken place within the first 100 Myr after CAIs~\citep{morby18,Nesvorny18b,Mojzsis19}, and some hint that it may have taken place within a few million years of the dispersal of the gaseous disk~\citep{Hunt2022,Edwards2024}.  From a dynamical point of view, the instability timing depends on the trigger, which is itself uncertain~\citep[although several mechanisms have been proposed; see][]{Gomes2005,Levison2011,RibeirodeSousa2020,Liu2022}.  Our choice of 20 Myr for the instability timing is intermediate -- early enough to potentially affect Mars' accretion~\citep{Clement2018}, but late enough not to assume the earliest possible trigger.
 
%%%%%%%%%%%%%%%%%%%%%%%%%%%%%%%%%%%%%%%%%%%%%%%%%%%%%%%%%%%%%%%%%%%%%%%%%%%%%%%%%%%%%%%%%%%%%%%%%%%%%%%%%%%%%%%%%%%%%%%%%%%%%%%%%%%%%%%%%%%
%%%%%%%%%%%%%%%%%%%%%%%%%%%%%%%%%%%%%%%%%%%%%%%%%%%%%%%%%%%%%%%%%%%%%%%%%%%%%%%%%%%%%%%%%%%%%%%%%%%%%%%%%%%%%%%%%%%%%%%%%%%%%%%%%%%%%%%%%%%
%%%%%%%%%%%%%%%%%%%%%%%%%%%%%%%%%%%%%%%%%%%%%%%%%%%%%%%%%%%%%%%%%%%%%%%%%%%%%%%%%%%%%%%%%%%%%%%%%%%%%%%%%%%%%%%%%%%%%%%%%%%%%%%%%%%%%%%%%%%
%%%%%%%%%%%%%%%%%%%%%%%%%%%%%%%%%%%%%%%%%%%%%%%%%%%%%%%%%%%%%%%%%%%%%%%%%%%%%%%%%%%%%%%%%%%%%%%%%%%%%%%%%%%%%%%%%%%%%%%%%%%%%%%%%%%%%%%%%%%
\section{Simulation Outcomes}%%%%%%%%%%%%%%%%%%%%%%%%%%%%%%%%%%%%%%%%%%%%%%%%%%%%%%%%%%%%%%%%%%%%%%%%%%%%%%%%%%%%%%%%%%%%%%%%%%%%%%%%%%%%%%%%%%%%%%%%%%
\label{cha:Results}

In this section, we first (Sec 3.1) describe the constraints we use to determine whether a simulation should be considered a viable analog to the Solar System.  Next (Sec 3.2), we present the detailed dynamical and collisional evolution of one simulation from the \emph{mixed} scenario is shown, both with and without the giant planet instability.  In Section 3.3, we explore the amount of CC mass accreted by terrestrial planet analogs all of the sets of simulations.  Then, we determine the CC mass fraction within the late accretion of these planets. Finally (Sec 3.4), we determine the compositions of the last giant impactors (Theia analogs) in all viable simulations, in the context of cosmochemical and dynamical constraints. 

Our inventory of simulations is as follows. We ran 60 simulations of the \emph{mixed} scenario, 10 of the \emph{large only} and 9 of the \emph{small only} scenarios.  We ran an additional 10 simulations (9 for the \emph{small only} set) also including the giant planet dynamical instability. 
 
\begin{figure}[!htb]
    \centering
    \includegraphics[width=0.49\linewidth]{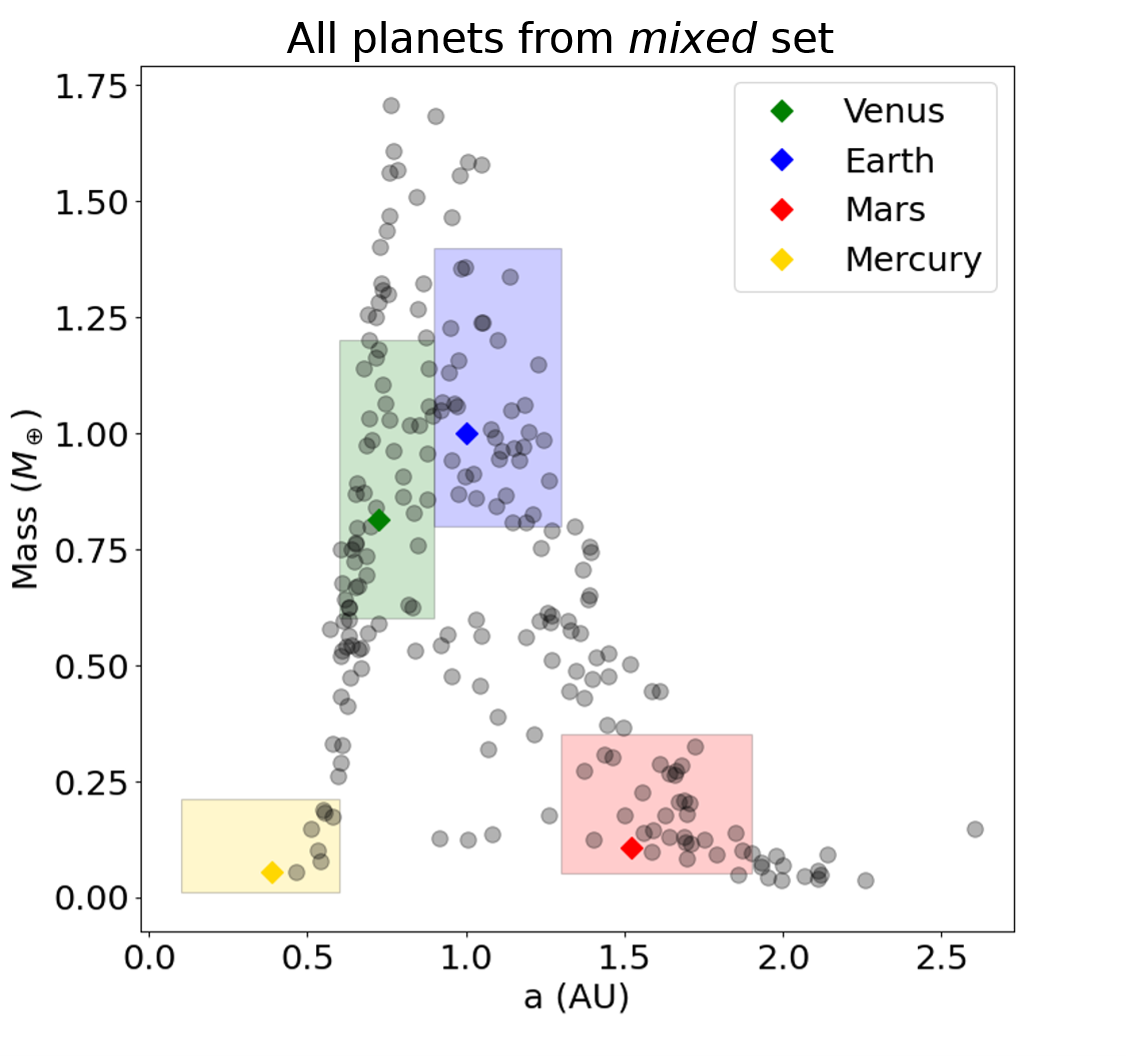}
        \includegraphics[width=0.49\linewidth]{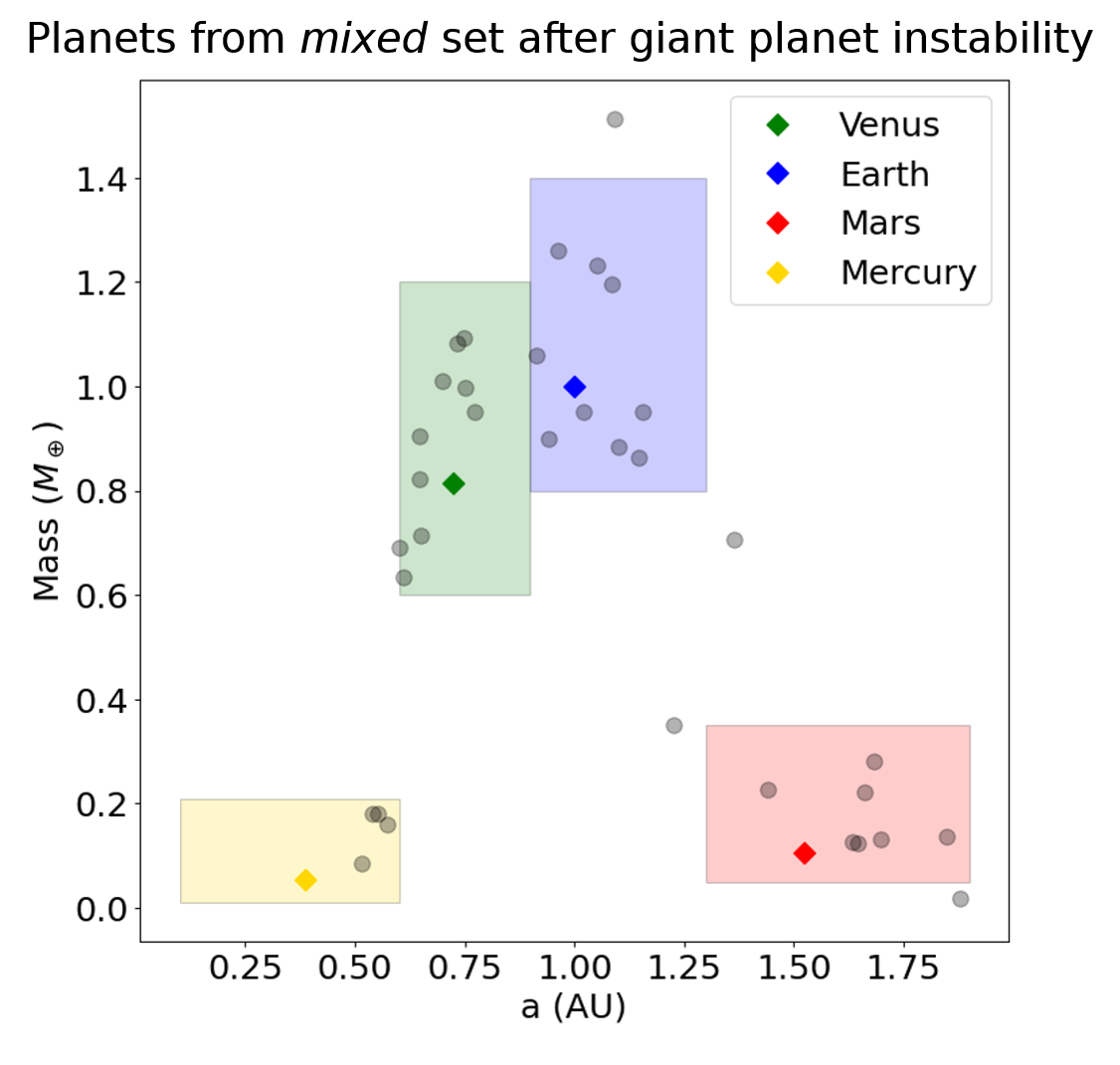}
    \caption{{\bf Left:} Mass and orbital radii of surviving large NC bodies after 200 Myr in all 60 simulations of the \emph{mixed} scenario. Constraints for planet selection are visualised with the boxes, and the real solar system planets are shown for comparison.  {\bf Right:} The final configuration in the 10 \emph{mixed} simulations that also included the giant planet instability.} 
    \label{fig:M_a_SnB}
\end{figure}

Figure \ref*{fig:M_a_SnB} shows the masses and orbital radii of the planets that formed in the \emph{mixed} scenario.  The boxes present the contours of our criteria for choosing analogs to each terrestrial planet.  They were chosen to be within $\sim$ a factor of 1.5 of the current values, with less strict constraints around Mars and Mercury analogues. Among simulations without the giant planet dynamical instability (left panel of Fig.~\ref*{fig:M_a_SnB}), the most common final arrangement was the formation of 3 to 4 terrestrial planets and in some rare occurrences 2 or 5 remained at the end. The formation of 4 terrestrial planets did not always mean that a Mercury analog formed, but rather that another, typically Mars-sized, planet formed between the Earth and Venus analogues or the Earth and Mars analogues. Indeed, such low-mass planets are common in the distribution from Fig.~\ref*{fig:M_a_SnB} (left panel). Including the giant planet instability helps remove those remaining Mars sized bodies, by forcing them to collide with the Earth or Venus analogs (see right panel of Fig.~\ref*{fig:M_a_SnB}).  Most other remaining NC and CC bodies, that were not the planet analogues, were also cleared out by the giant planet instability.  In most cases all CC embryos were accreted or ejected from the solar system. However, a rare occurrence for both the \emph{large only} and \emph{mixed} scenarios was the survival of one CC embryo beyond the orbit of Mars.

\begin{figure}[ht]
    \centering
    \includegraphics[width=0.49\linewidth]{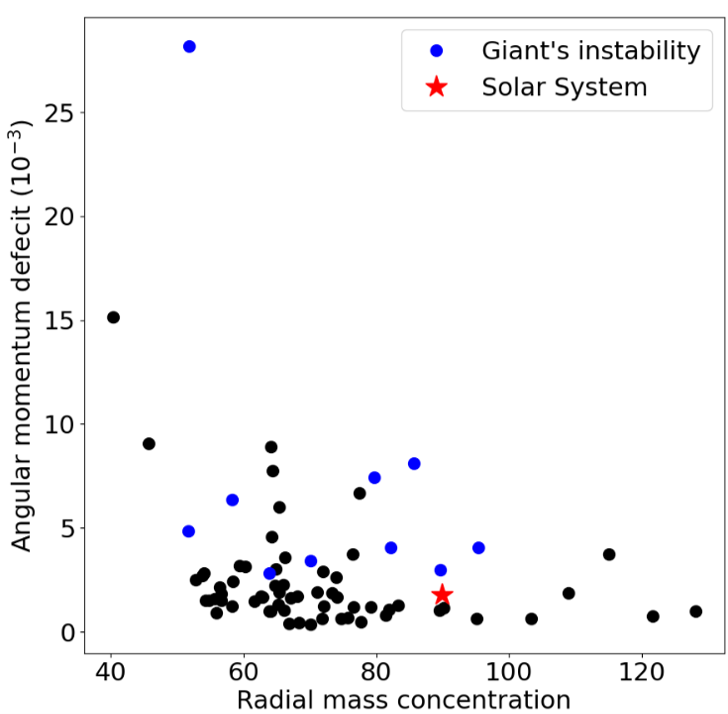}
    \caption{Normalized angular momentum deficit $AMD$ and radial mass concentration $RMC$ for the 60 simulations in using the \emph{mixed} scenario (black) and the 10 giant planet instability ones (blue), as well as a comparison with the real solar system value (red).}
    \label{fig:RMC_AMD_SnB}
\end{figure}

Figure \ref*{fig:RMC_AMD_SnB} compares the radial mass concentration $RMC$ and normalized angular momentum deficit $AMD$ of the terrestrial planets in the \emph{mixed} scenario with and without the giant planet instability with the actual terrestrial planets. 
The instability seems to increase the angular momentum deficit of the simulations, while keeping the radial mass concentration similar. 

\FloatBarrier

\subsection{Evolution of an individual simulation} 
In this subsection we present the evolution of a promising Solar System analog simulation from the \emph{mixed} scenario.  We then present how the evolution changed when the giant planet instability was included.

\begin{figure}[!htb]
    \centering
    \includegraphics[width=0.8\linewidth]{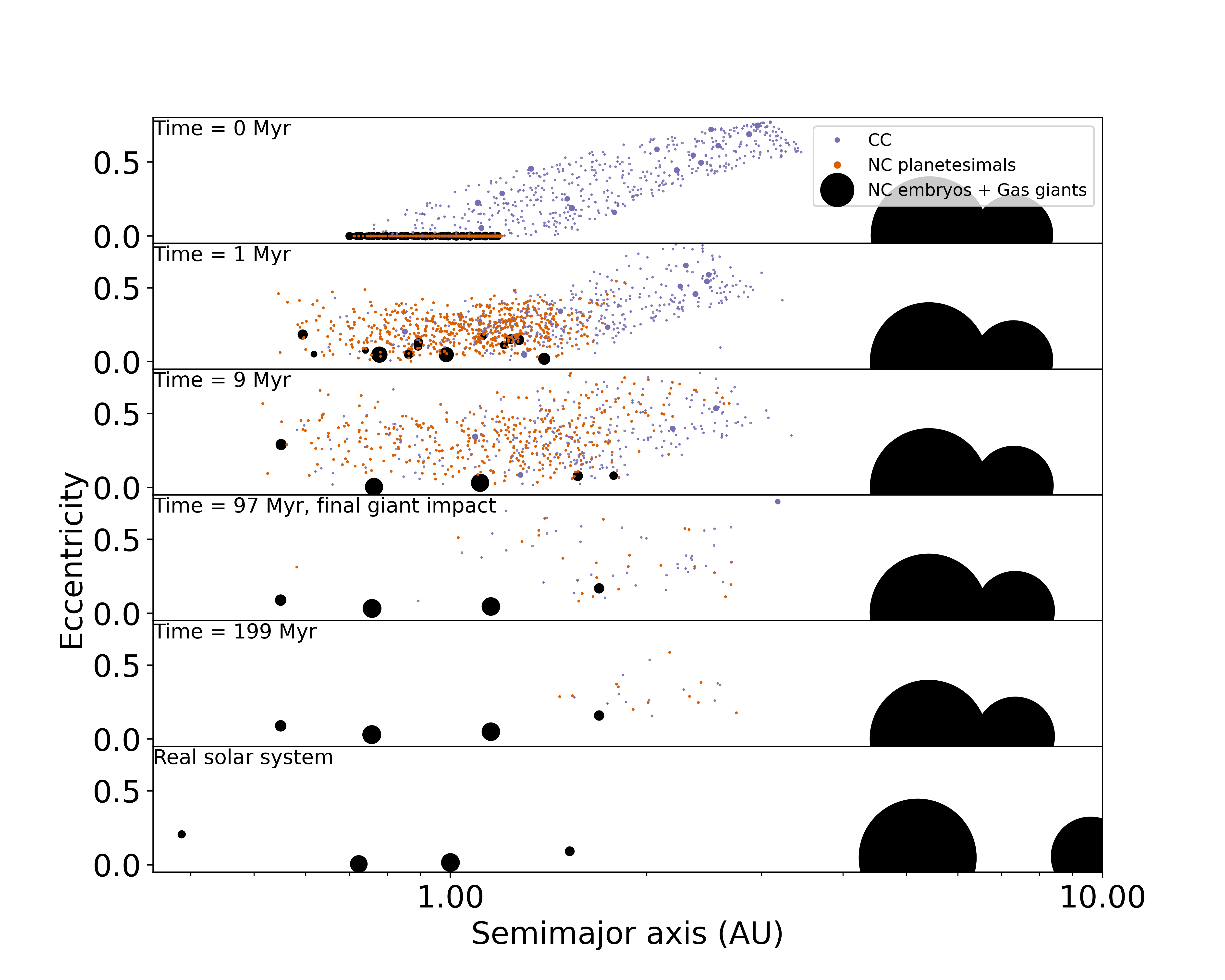}
    \caption{Eccentricity and position snapshots over time of a simulation using the \emph{mixed} scenario where all 4 terrestrial planets fit constraints and the Earth analogue had a final CC embryo impact. Snapshots are shown over time, including the moment after the final giant impact. The initial NC embryos and planetesimals are shown with the embryos in black and the planetesimals in orange, as well as the CC embryos and planetesimals in blue with Jupiter and Saturn in black on the far right. The final snapshot is the real solar system.}
    \label{fig:e_a_SnB}
\end{figure}

\begin{figure}[!htb]
    \centering
    \includegraphics[width=0.6\linewidth]{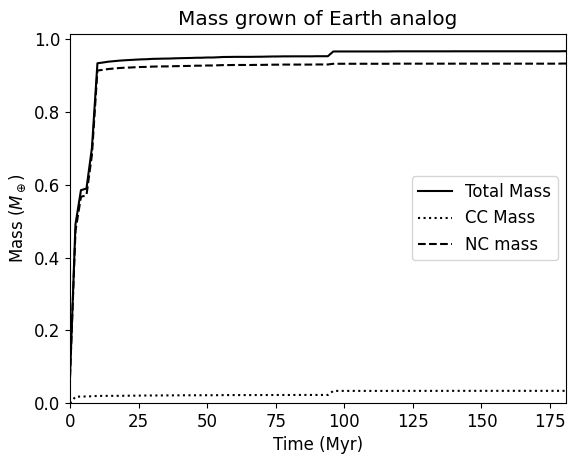}
    \caption{Growth curve of Earth analogue in the simulation from figure \ref{fig:e_a_SnB}. Total mass is separated into contribution from NC and CC bodies.}
    \label{fig:Growth_Curve_20}
\end{figure}

Figure~\ref*{fig:e_a_SnB} shows the evolution of the simulation, without including the giant planet dynamical instability.  At early times, the scattered CC objects mix with the NC bodies in the terrestrial planet-forming region, while leaving some large CCs orbiting in between planets or still too far to collide. The simulation ended with 4 terrestrial planets, including a reasonable Earth analog with 1 M$_\oplus$ and Mars analog with 0.1 M$_\oplus$.  The $RMC$ and $AMD$ for this simulation are 66 and 2.27$\times 10^{-3}$ respectively. The final giant impact on the Earth analog happened with a CC embryo at 97 Myr. Figure~\ref*{fig:Growth_Curve_20} shows the growth curve of the Earth analogue separated into the NC and CC contributions. The final giant impactor's contribution can be seen in the sudden late and large gain of CC mass before 100 Myr.

\begin{figure}[!htb]
    \centering
    \includegraphics[width=0.8\linewidth]{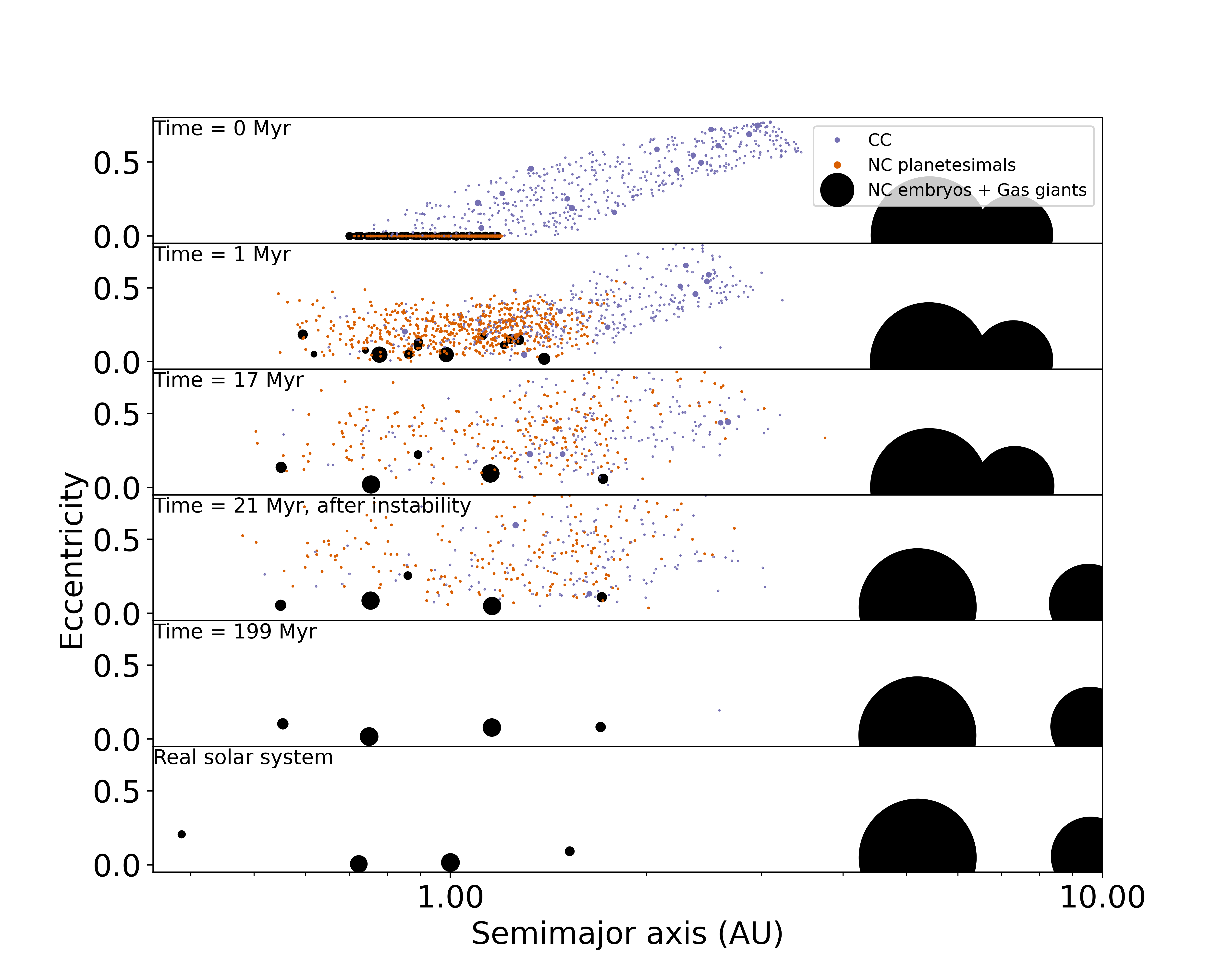}
    \caption{Eccentricity and position snapshots over time of the simulation in figure \ref*{fig:e_a_SnB} now with the giant planet instability at 20 Myr. The initial NC embryos and planetesimals are shown with the embryos in black and the planetesimals in orange, as well as the CC embryos and planetesimals in blue with Jupiter and Saturn in black on the far right. The final snapshot is the real solar system.}
    \label{fig:e_a_SnB_GPI}
\end{figure}

Figure \ref*{fig:e_a_SnB_GPI} shows the evolution of the exact same simulation when the giant planet instability is taken into account, as Jupiter and Saturn's orbits are instantaneously moved to their present-day ones at 20 Myr.  The giant planet instability dramatically changed the evolution of the system causing a strong pulse of eccentricity excitement, which lead to a wave of collisions and ejections.  The final state of the system is only modestly-changed from the instability-free case from Fig.~\ref*{fig:e_a_SnB} -- there are still four terrestrial planets with the Earth analogue having 1 M$_\oplus$ and the Mars analogue having 0.1 M$_\oplus$.  The $RMC$ and $AMD$ for this simulation are 63.9 and 2.81$\times 10^{-3}$, respectively.  However, the final giant impact on the Earth analog was quite different, happening at 10 Myr with a dry, NC embryo.

%\begin{figure}[!htb]
%    \centering
%    \includegraphics[width=0.6\linewidth]{Ch3/aqQ_SnB.png}
%    \caption{Semi-major axis $a$, aphelion $Q$ and perihelion $q$ evolution for objects inside the \emph{mixed} scenario simulation from figure \ref*{fig:e_a_SnB}. The final impactor is a CC marked in red that collided with the Earth analogue, the black line around 1 AU.}
%    \label{fig:aQq_SnB}
%\end{figure}

\FloatBarrier
\subsection{The CC mass fraction} %FULL CHECK
\label{sec:CCmass}

A key constraint on our simulated planets is the fraction of their mass that was accreted from carbonaceous material, as well as the timing of its accretion. Cosmochemical studies indicate that the CC fraction for Earth is a few to 10-15\%~\citep[e.g.][]{dauphas17,martins23,KLEINE2023,Dauphas2024,Nimmo2024}.  \cite{Nimmo2024} and \cite{Dauphas2024} argued that that at least some of this CC material was delivered during the last $\sim 10\%$ of our planet's accretion. For Mars, the CC fraction is lower, between almost zero and $2-3 \%$~\citep{KLEINE2023}. There are no direct measurements of the other terrestrial planets due to the lack of samples. 

\begin{figure}[htbp]
    \centering
    \includegraphics[width=0.8\linewidth]{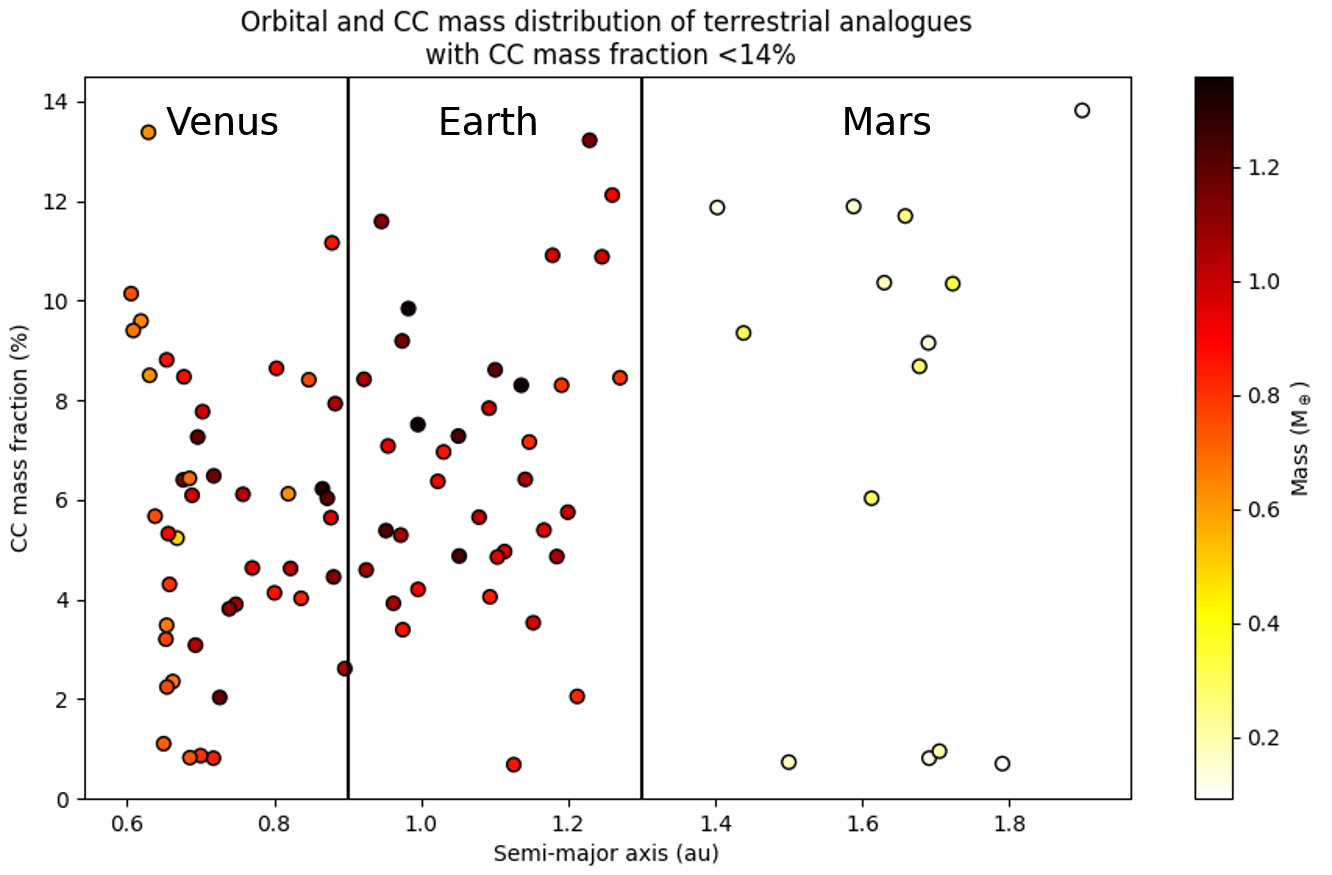}
    \caption{Orbital and CC mass distribution of Earth, Venus and Mars analogues in the \emph{mixed} sets with CC mass fractions under 14\%. Colours represent the mass of the analogue in Earth masses.}
    \label{fig:e_a_CC_All}
\end{figure}

Figure \ref*{fig:e_a_CC_All} shows the orbits and CC mass fractions of Venus, Earth and Mars analogues and their masses for the \emph{mixed} simulations. The Earth analogs are a good match, with a median [mean] of 6.4\% [6.8\%] in \emph{mixed} simulations with no instability, and 5.1\% [5.3\%] in simulations with a giant planet dynamical instability. Venus analogs tend to have a similar CC mass fraction as Earth analogs, with typical values of $\sim 5\%$. 

In the \emph{small only} simulations, the typical CC fraction of Earth is somewhat lower, with a median [mean] of 4.1 \% [4\%] in \emph{mixed} simulations with no instability, and 3.6\% [3.6\%] in simulations with a giant planet dynamical instability.  Venus analogs in the \emph{small only} simulations have typical CC mass fractions of $\sim 3-4\%$.  In contrast, the CC mass fractions are higher in the \emph{big only} simulations.  The median [mean] CC mass fraction of Earth analogs is 8.8\% [8.4\%] in simulations with no giant instability, and 8.1\% [8.9\%] in simulations with a giant planet instability.  Venus had a somewhat higher CC mass fraction than Earth in the \emph{big only} simulations, with a median [mean] of 8.7\% [12.1\%] in \emph{mixed} simulations with no instability, and 8\% [9.4\%] in simulations with a giant planet dynamical instability. The giant planet instability did not seem to affect the CC mass provided by the final impactor in the simulations. 

Within the \emph{small only} scenario, Mars and Earth analogues each accreted with a similar CC mass fraction, typically around 4\%. This was likewise found in the simulations of \cite{JOIRET2024}.  This served as motivation for the idea presented by \cite{Nimmo2024} and \cite{JOIRET2024} that CC objects in the terrestrial realm must have included larger objects (planetary embryos).

Let us now focus on the \emph{mixed} scenario simulations, specifically on simulations in which the Earth/Mars mass ratio was within a factor of two of its current value.  Mars analogs followed one of two paths: they were either hit by a CC embryo or not. In 62.5\% of simulations, the Mars analogue was hit by a CC embryo, and due to its size, the CC mass fraction of the analogue jumped above 10\%. When the analogue was not hit, its CC mass fraction was between 0.5\% and 1\%. In these simulations Earth analogues had an average CC mass fraction of $\sim$6\%, consistent with cosmochemical constraints for  Earth and Mars~\citep{KLEINE2023,Nimmo2024}.  These are simulations with no giant planet instability.  If we include the instability, Mars systematically receives less CC material. In the \emph{mixed} scenario, the Mars analogue is hit 75\% of the time leading still to a CC mass fraction above 10\% and when not hit the fraction is under 0.5\% instead.
In the \emph{small only} scenario, both Mars and Earth analogues again finish with about a 3.6\% CC mass fraction. 

\FloatBarrier

\subsection{The last giant impact} %FULL CHECK
\label{sec:finalgiantcollision3}

We now address the details of the last giant impacts in our simulations, which are the potential Moon-forming collisions~\citep[e.g.][]{Canup2001}.  We focus specifically on the origin of the impactor -- Theia -- in these last collisions, as \cite{Budde2019} claimed that it may have been a carbonaceous object.

\begin{figure}[htbp]
    \centering
    \includegraphics[width=0.49\linewidth]{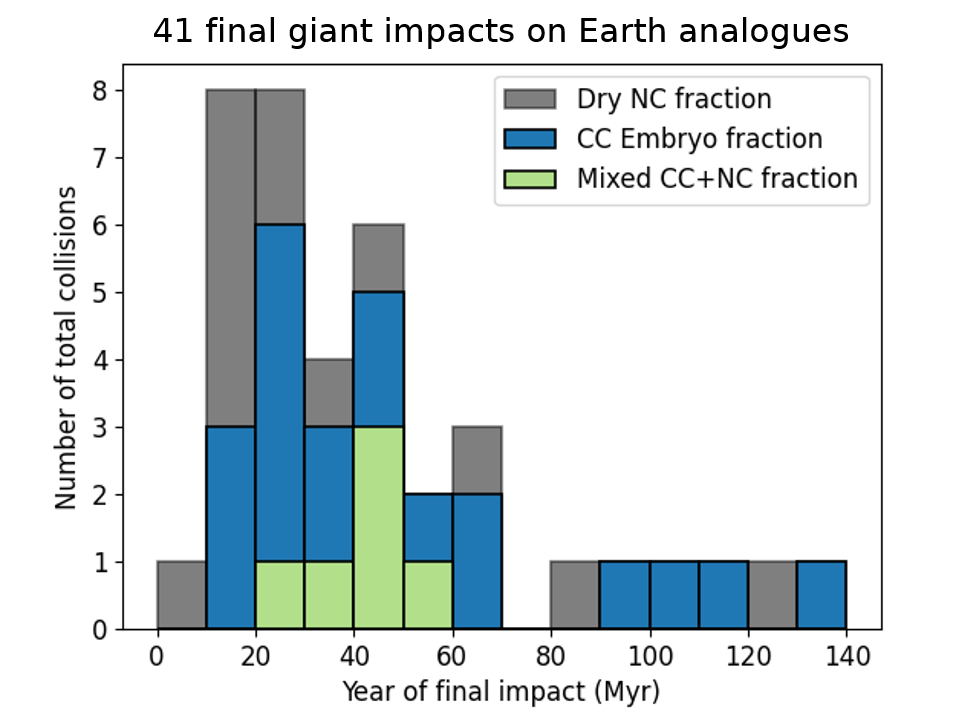}
        \includegraphics[width=0.49\linewidth]{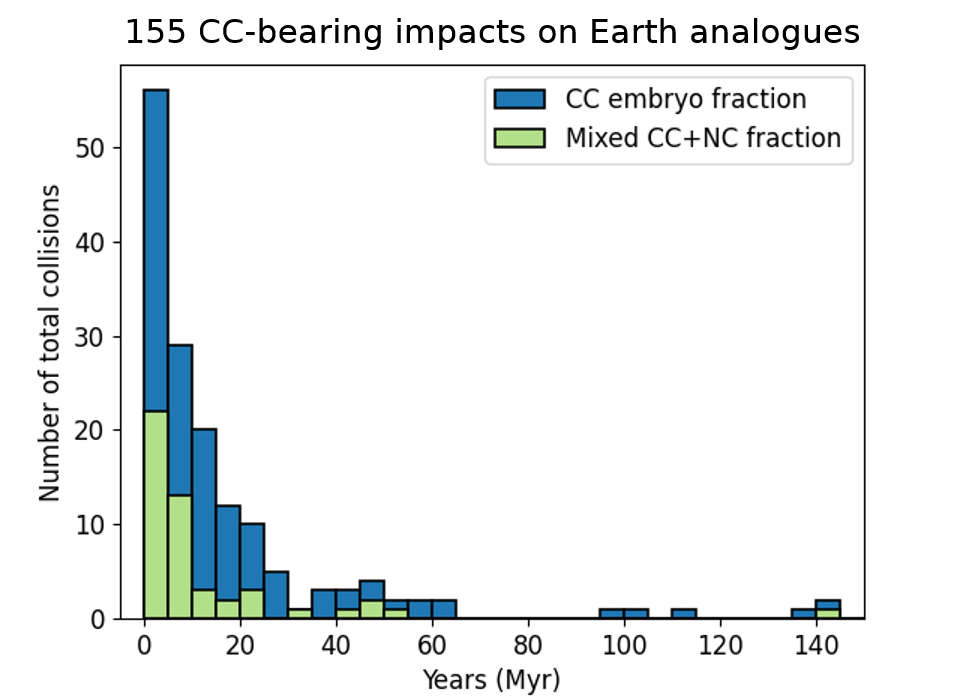}
    \caption{{\bf Left:} Distribution of 41 final giant impacts with Earth analogues. The blue bars represent the CC embryo final impactor fraction, the green bars are the mixed NC+CC embryo (second order) fraction, while the gray bars represent dry NC embryo fraction.  {\bf Right}: Distribution in time of 155 giant impacts with CC components with Earth analogues. Blue bars represent the CC embryo impactor fraction and the green bars are the mixed NC+CC embryos (second order) fraction.}
    \label{fig:Col_Hist_SnB}
\end{figure}

In the \emph{mixed} scenario with no giant planet instability, Earth's final impactor included a CC component in more than half of all simulations. In 38.5\% of simulations, the final impactor was a pure CC embryo, and in another 13.5\%, the impactor was an NC embryo that had previously accreted a CC embryo. Figure \ref*{fig:Col_Hist_SnB} (left panel) shows the distribution of the impact timing for last giant impacts, broken down by the composition of the impactor. Figure \ref*{fig:Col_Hist_SnB} (right panel) shows the distribution of all giant impacts with Earth analogues that contained a CC component.  The mixed fraction decays much faster than the CC embryo fraction with most impacts happening before 10 Myr. This is because, once a CC embryo has collided with an NC embryo, the new larger embryo is by default more centrally located, likely within the heart of the annulus, and likely to undergo additional collisions on a short timescale.

\begin{figure}[htbp]
    \centering
    \includegraphics[width=0.7\linewidth]{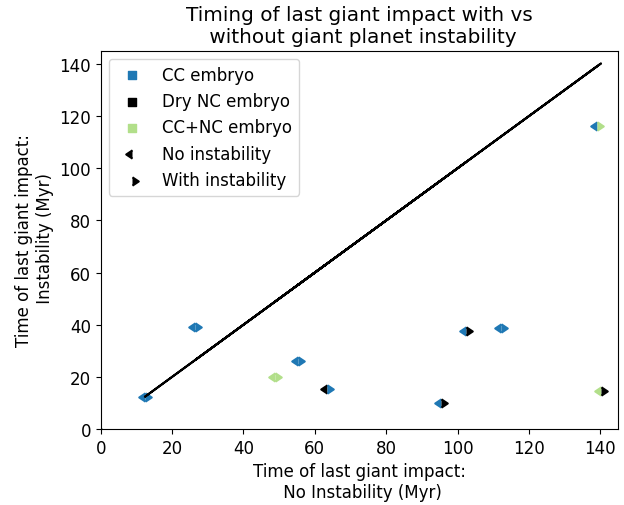}
    \caption{A comparison between the timing of the last giant impacts in 10 \emph{mixed} simulations simulations that were run both with and without the giant planet instability. The black line represents the point where both values are equal. Each point has two halves with the left half representing the impactor type in the simulation without the giant planet instability and the right half representing the simulation with the giant planet instability. Dry NC impactors are black, CC embryos are blue and CC+NC mixed embryos are green.}
    \label{fig:GPI_Impact_Relations}
\end{figure}

Figure \ref*{fig:GPI_Impact_Relations} shows the relative timing of the last giant impact for the ten simulations from the \emph{mixed} set that were run both with and without the giant planet dynamical instability.  The punchline is simply, once again, that including the giant planet dynamical instability leads to earlier last giant impacts on Earth analogs.  This is because the instability tends to excite terrestrial embryos and stimulate giant impacts.  Our results are consistent with those of \cite{DeSouza2021}, who found a typical delay of $\sim 20$~Myr between the time of the instability and the last giant impact.

\begin{figure}[htbp]
    \centering
    \includegraphics[width=0.7\linewidth]{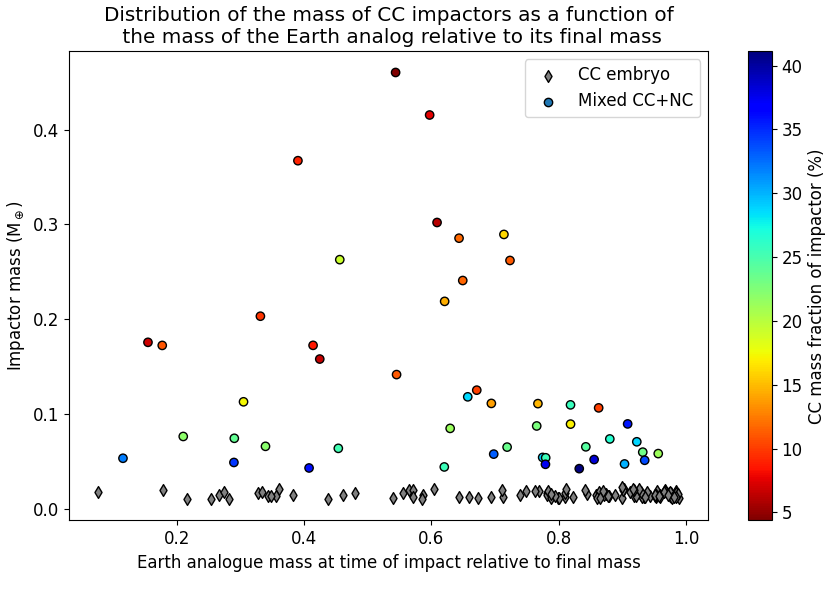}
        \includegraphics[width=0.8\linewidth]{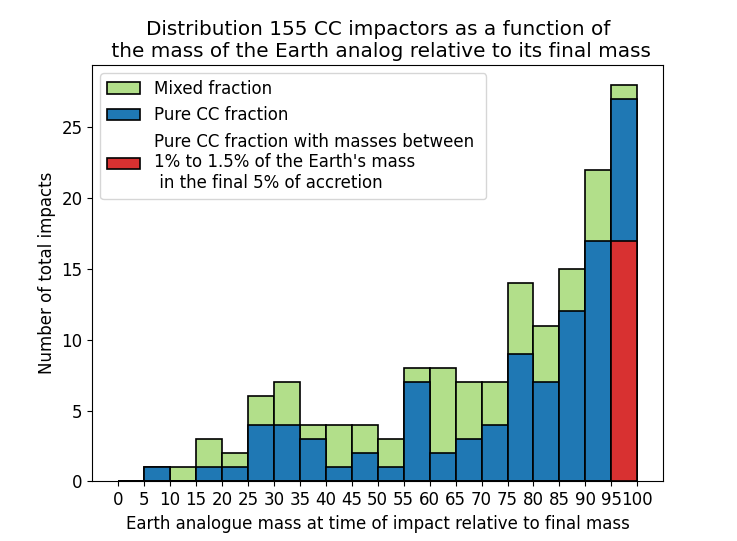}
    \caption{{\bf Top:} Mass distribution of CC impactors on Earth analogues in the \emph{mixed} sets, in relation to the analogue's mass at time of impact relative to its final mass. Diamonds are pure CC impactors, while circles are mixed CC+NC impactors with their colour representing their CC mass fraction. {\bf Bottom:} Distribution of 155 CC bearing impacts in relation to the analogue's mass at time of impact relative to its final mass. Green fraction represents mixed NC+CC impactors while blue represents the pure CC fraction. In red is the fraction of pure CC objects with masses between 1\% to 1.5\% of the Earth's mass which impacted in the final 5\% of accretion.}
    \label{fig:Col_Hist_Mass}
\end{figure}
\FloatBarrier

\subsection{Late accretion} %FULL CHECK
\label{sec:lateaccretion}

We now turn our attention to the CC mass fraction of the late accretion phase.  Recall that late accretion represents planetesimal impacts after the last giant (embryo) impact~\citep[see, e.g.][]{Jacobson2014,MorbyWood2015}.  There is cosmochemical evidence that late accretion was dominated by dry (NC) material~\citep{Gillmann2020,Worsham2021}.

\begin{figure}[htbp]
    \centering
    \includegraphics[width=0.7\linewidth]{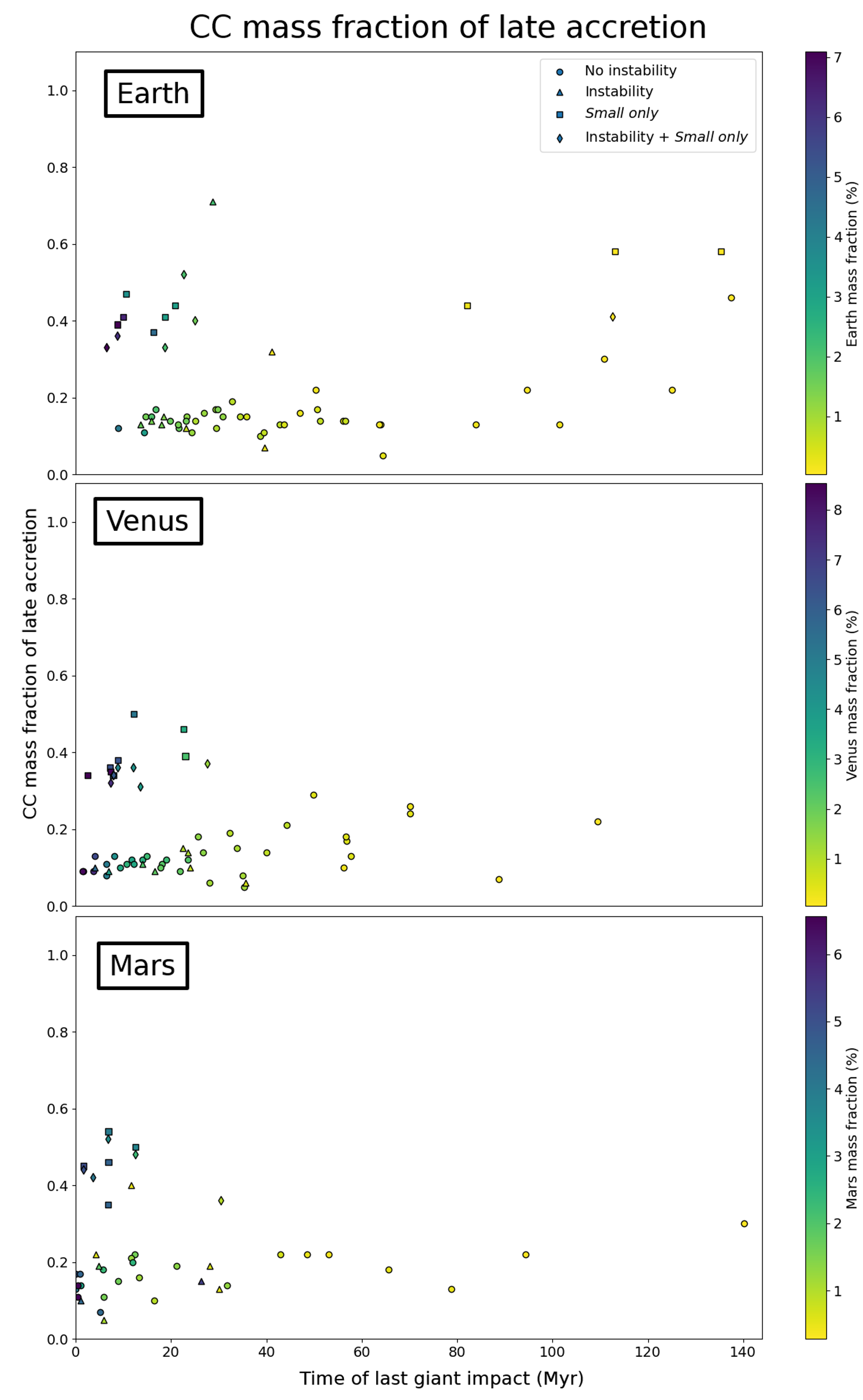}
    \caption{CC mass fraction of late accretion for Earth, Venus and Mars analogues, as a function of the time of the last giant impact (which marks the start of late accretion). The circles and triangles are from the \emph{mixed} scenario. Squares and diamonds mark the \emph{small only} simulations. The colours indicate the total mass fraction accreted during late accretion.}
    \label{fig:LA_MR}
\end{figure}

Figure \ref*{fig:LA_MR} shows the carbonaceous (CC) mass fraction of late accretion for Earth, Venus and Mars analogs as a function of the time of the last giant impact (i.e., the start time of late accretion). It is immediately clear that simulated planets in the \emph{small only} scenario have a much higher CC fraction of late accretion than in the \emph{mixed} scenario.  This appears to be a simple consequence of the fact that the \emph{small only} simulations have far more total mass in CC planetesimals, and enough CC planetesimals survive to provide a significant amount of late accretion~\citep[this was also found by][]{JOIRET2024}.

The \emph{mixed} scenario has a much more NC-rich late accretion phase in general (Fig.~\ref*{fig:LA_MR}).  The majority of CC material is delivered to each of the terrestrial planets during the main phase of accretion. In most cases, Venus has a late accretion with only a small fraction of CC material, broadly consistent with the results of  \cite{Gillmann2020}.  It is noticeable that the CC fraction rises modestly when the last giant impact happens late.  This is a byproduct of the fact that a later last giant impact correlates with less late accretion~\citep{Jacobson2014} -- since CC planetesimals are modestly farther from the Sun than NC ones on average, their collision rates are lower and a higher fraction survive to late times. 

The effect of the giant planet dynamical instability timing is also evident in Fig.~\ref*{fig:LA_MR}.  None of the simulations with the giant planet instability has a last giant impact later than $\sim 40$ Myr on Earth.  This is a simple byproduct of the eccentricity excitation induced by the instability, which led to a rapid clearing of stray embryos by collisions.  In a handful of cases, the dynamical instability also resulted in an elevated CC fraction for late accretion in Earth and Mars analogs. 

\FloatBarrier

%%%%%%%%%%%%%%%%%%%%%%%%%%%%%%%%%%%%%%%%%%%%%%%%%%%%%%%%%%%%%%%%%%%%%%%%%%%%%%%%%%%%%%%%%%%%%%%%%%%%%%%%%%%%%%%%%%%%%%%%%%%%%%%%%%%%%%%%%%%
%%%%%%%%%%%%%%%%%%%%%%%%%%%%%%%%%%%%%%%%%%%%%%%%%%%%%%%%%%%%%%%%%%%%%%%%%%%%%%%%%%%%%%%%%%%%%%%%%%%%%%%%%%%%%%%%%%%%%%%%%%%%%%%%%%%%%%%%%%%
%%%%%%%%%%%%%%%%%%%%%%%%%%%%%%%%%%%%%%%%%%%%%%%%%%%%%%%%%%%%%%%%%%%%%%%%%%%%%%%%%%%%%%%%%%%%%%%%%%%%%%%%%%%%%%%%%%%%%%%%%%%%%%%%%%%%%%%%%%%
%%%%%%%%%%%%%%%%%%%%%%%%%%%%%%%%%%%%%%%%%%%%%%%%%%%%%%%%%%%%%%%%%%%%%%%%%%%%%%%%%%%%%%%%%%%%%%%%%%%%%%%%%%%%%%%%%%%%%%%%%%%%%%%%%%%%%%%%%%%
\section{Summary and Discussion}%%%%%%%%%%%%%%%%%%%%%%%%%%%%%%%%%%%%%%%%%%%%%%%%%%%%%%%%%%%%%%%%%%%%%%%%%%%%%%%%%%%%%%%%%%%%%%%%%%%%%%%%%%%%%%%%%%%%%%%%%%%%%%%%%%%%%%%%%%%
\label{cha:Discussion}

Our primary result is to show that it is plausible that the last giant impactor on Earth (Theia) was a carbonaceous (CC) object, in the context of a full dynamical scenario for the accretion and early evolution of the terrestrial planets. The basic idea that Theia may have been carbonaceous was proposed by \cite{Budde2019}, and the idea that carbonaceous material in the inner Solar System included embryos was proposed by \cite{Nimmo2024} and \cite{JOIRET2024}. %corrected joiret tag  

The scenario is illustrated in our \emph{mixed} simulations and proceeds as follows.  First, planetesimal formation started quickly in the Sun's protoplanetary disk, likely at locations associated with condensation fronts~\citep{Armitage2016,drazkowska18}.  This led to two or more rings of planetesimals -- the inner one was rocky, non-carbonaceous (NC), and centered around 1 au, containing a total of $\sim 2-2.5 \mearth$~\citep{Morbidelli2022,Izidoro2022}.  The giant planets' cores grew from a more distant ring at $\sim 5-10$ au.  The rapid gas accretion phase of the gas giants, as well as the migration of the ice giants, scattered CC material inward.  Some planetesimals were trapped in the asteroid belt (as the C-types) and, given their weaker gas drag, preferentially larger bodies were scattered onto terrestrial planet-crossing orbits~\citep{RaymondIzidoro2017}.  The late-stage accretion of the terrestrial planets involved a series of giant impacts between NC embryos and planetesimals, with occasional impacts of CC objects.  The last giant impact on the proto-Earth produced the Moon~\citep[e.g.][]{Canup2001} and marked the start of late accretion phase of planetesimal impacts. The giant planet dynamical instability took place during the main phase of terrestrial accretion~\citep{morby18,Nesvorny18b,Mojzsis2019}, likely within $\sim 20$~Myr of gas disk dispersal~\citep{RibeirodeSousa2020,Liu2022,Hunt2022,Edwards2024}, although the exact timing remains a matter of debate. 

This scenario can explain a number of Solar System constraints.  First, the general setup can adequately match the masses and orbits of the terrestrial planets, including the large Earth/Mars mass ratio~\citep[as shown by][]{Hansen_2009}. Our simulations also provide a decent match to the Solar System's normalized Angular Momentum Deficit $AMD$ and Radial Mass Concentration $RMC$. Second, the scenario can match the orbital distribution of the asteroids, as NC planetesimals (largely S-types) were scattered outward from the terrestrial planet-forming region~\citep{RaymondIzidoro2017b,Izidoro2024} and C-types were scattered inward from the giant planet region~\citep{RaymondIzidoro2017}. Our simulations do not model this phase. 

Third, this scenario can match the CC mass fraction of Earth and Mars.  Earth's CC mass fraction is constrained by measurements to be a few to $10-15\%$, with typically inferred values of $\sim 6\%$~\citep{KLEINE2023,Dauphas2024,Nimmo2024}.  In contrast, Mars' CC mass fraction is much lower, at most a few percent~\citep{KLEINE2023}. Our simulations tend to form Earth analogs with CC mass fractions of a few to ten percent, in agreement with this constraint.  If the CC material was entirely in the form of planetesimals (the \emph{small only} simulations), the mass fraction of Earth and Mars would be roughly the same~\citep[as also shown by][]{JOIRET2024}. However, if the mass in CCs is predominantly in embryos, our simulations can explain Earth's much larger CC mass fraction as long as Mars was not hit by a CC embryo~\citep[as proposed by ][]{Nimmo2024}.  In 62.5\% of simulations, the Mars analog is indeed hit by a CC embryo, causing its CC mass fraction to jump above 10\%.  Such outcomes are not consistent with cosmochemical measurements~\citep{KLEINE2023}. In simulations where Mars analogs are not hit by a CC embryo, their CC mass fraction is typically $\sim 0.5-1 \%$ and Earth's is $\sim 5-7\%$, both of which match constraints. 

Fourth, this scenario can match the approximate timing of the Moon-forming impact. The last embryo impact took place in our simulations after 5-150 Myr, with a large fraction of last giant impacts at 20-70 Myr. Given the uncertainties in interpreting the timing of Moon formation~\citep[e.g.][]{Fischer2014,Nimmo2024b}, our simulations produce viable realizations.  

Finally, this scenario can produce a volatile-poor, NC-dominated late accretion phase. Late accretion is constrained by models and cosmochemical measurements to have been NC-dominated~\citep{Gillmann2020,Worsham2021}.  The late accretion phase in our \emph{mixed} simulations is indeed NC-dominated, for the simple reason that only a small fraction of the CC mass was in planetesimals. Mars analogues also had a higher median of CC/NC mass fraction accreted during late accretion than Earth and Venus analogues, with Venus having the lowest.

Within the context of this scenario, the last giant impactor on Earth contained a CC component in roughly half of all of the \emph{mixed} simulations.  In the majority of these (38\% of simulations), Theia was a pristine CC embryo, and in the remainder of cases Theia was an NC embryo that had previously accreted a CC embryo. This offers credence to the idea of \cite{Budde2019}.  As modeled by \cite{Nimmo2024}, CC embryos and planetesimals can be accreted throughout Earth's growth, but they are concentrated toward the later phases. Figure~\ref*{fig:Col_Hist_Mass} supports this with most impacts from embryos with CC components happening after the Earth analogue has accreted 80\% of its final mass, with a considerable fraction of lower mass pure CC embryos impacting after 95\% of accretion.

It is worth keeping in mind that \cite{Bermingham2025} came to the opposite conclusion: that the last 10-20\% of Earth's accretion was likely NC-rich.  In this scenario, Earth's CC budget must have been accreted in the first 80-90\% of accretion. This is also entirely plausible within the context of our simulations (see Figure~\ref*{fig:Col_Hist_Mass}), as many CC-rich embryo impacts take place in that window.  Evaluating the exact timing of CC accretion appears to require cosmochemical analysis, as our simulations cannot distinguish between the scenarios of \cite{Nimmo2024} and that of \cite{Bermingham2025}.

Using our results as a guide, we can estimate the initial mass in CC planetesimals relative to CC embryos as follows.  In the \emph{small only} simulations, Earth and Mars tended to have the same CC mass fraction of $\sim 4\%$ (as did Venus). Let's assume that Earth and Mars generally accrete the same fraction of the planetesimal mass, as normalized by their mass ratio (of $\sim 9$). Let's also take the results of \cite{KLEINE2023} at face value, and adopt a Mars total CC mass fraction of $\sim 0.5\%$ and a value of $\sim 6\%$ for Earth. This means that $\sim 0.5\%$ of Earth's mass was delivered by CC planetesimals, and the remaining $\sim 5.5\%$ by CC embryos. From the \emph{mixed} set of simulations, Earth analogues accreted a median of 5.7\% of CC material from embryos and 0.7\% from planetesimals. Mars analogues also accreted a median of 0.7\% CC mass from CC planetesimals, which follows the logic of the approximation made before. There must therefore have been $\sim 8$ times more mass in CC embryos than in CC planetesimals.

We can also use simple arguments to evaluate the typical CC embryo mass.  If there were too many embryos, it would be extremely unlikely that Mars could have avoided a CC impact.  The number of embryos for which the odds of Earth accreting embryos but Mars not accreting any should be roughly the same as the their accretion ratio, which is simply their mass ratio of $\sim 9$ (because they accreted the same mass fraction in CC planetesimals). Indeed, in our \emph{mixed} simulations with 15 embryos, Mars accreted a CC embryo in more than half (62.5\%) of simulations. To avoid dropping the odds of success any further, we do not expect CC embryos to have been less massive than the 1-2\% of an Earth-mass assumed here.  On the other hand, the most massive CC embryo cannot have been more massive than the $\sim 5.5\%$ of Earth's mass that was accreted in the form of CC embryos. From~\cite{Budde2019}, we also have a prediction of the CI material provided by the final impactor -- 2\%$\pm$1\%. Depending on the nature of the CC embryos, this can further limit the mass of the final impactor. A reasonable mass range for CC embryos is therefore $\sim 1-5\%$ of Earth's mass.  This is below the mass range for Theia in the canonical Moon-forming impact~\citep{Canup2001}, but could represent a plausible Moon-forming impact as long as the proto-Earth was spinning fast~\citep{Cuk2012}.  If a larger Theia mass were needed to match the Earth-Moon system~\citep{canup2021origin}, then the most viable option is an NC embryo that previously accreted a CC embryo -- a scenario that happened for 13.5\% of our Earth analogs in the \emph{mixed} set.

%which begs the question of whether a single CC impact could represent a viable Moon-forming event~\citep[or whether multiple such impacts would be needed, as proposed by ][]{Rufu2017}.  This is a reasonable question to pose in the context of the conjecture of \cite{Budde2019} as well as our own results. If a larger Theia mass is needed to match the Earth-Moon system~\citep{canup2021origin}, then the most viable option is an NC embryo that previously accreted a CC embryo -- that scenario happened on 13.5\% of our Earth analogs in the \emph{mixed} set. 

Put together, this reasoning implies that the total mass in CC material injected into the inner Solar System was $\sim 0.2-0.3 \mearth$, with a $\sim 8:1$ mass division between embryos and planetesimals.  The total mass ensures a match to Earth's CC mass fraction, whereas the embryo/planetesimal mass balance ensures that Mars did not accrete a CC embryo.  An even higher embryo-to-planetesimal CC mass ratio may be needed, to ensure that the late accretion phase was NC-dominated. 

If $\sim 0.2-0.3 \mearth$ of CC material was injected into the inner Solar System, it implies the existence of a much larger primordial reservoir in the outer Solar System.  The efficiency of implantation depends on the relative importance of giant planet growth and migration~\citep{Walsh2011,RaymondIzidoro2017}, as well as the strength of gas drag (connected with the object size).  The implantation efficiency is far higher for objects originating in the Jupiter-Saturn region (interior to $\sim 10$~au) than in the ice giant realm~\citep[out to $\gtrsim 20$~au;][]{RaymondIzidoro2017,Nesvorny2024}. Assuming an implantation efficiency of 1-10\%, this implies that 2-30 $\mearth$ in unaccreted CC embryos existed in the Jupiter-Saturn region.  If the gas giants' cores were $\sim 10 \mearth$ each, this suggests that the accretion efficiency of CC embryos was at most $\sim 90\%$ and perhaps as low as 50\% (or less).  This fits with the well-established paradigm that the growth of the giant planets' cores requires accretion of strongly damped objects (pebbles)~\citep{Ormel2010,Lambrechts2012}, as large planetesimals and embryos tend to gravitationally scatter each other rather than accrete~\citep[e.g.][]{Thommes2003}.

Our exploration of this scenario admittedly includes some uncertainties and limitations.  As our simulations started after gas disk dispersal, we did not model the injection of carbonaceous objects into the inner Solar System in a self-consistent fashion.  In addition, we made the assumption that an annulus of planetary embryos and planetesimals could be maintained to the end of the gas disk phase, even though migration and drag might make that impossible unless there was a convergent migration zone~\citep{Woo_2023}.  Finally, our implementation of the giant planet dynamical instability was highly simplified, although we note that it did broadly match the outcomes seen by \cite{DeSouza2021}.

\section*{Acknowledgement} 
We thank the anonymous referees for their constructive reports, and acknowledge helpful discussions with Katherine Bermingham.
We would like to thank the Faculdade de Ciências da Universidade de Lisboa for providing us with access to the cluster used for the simulations in this paper. 
We would also like to thank the Laboratoire d'Astrophysique de Bordeaux, the Université de Bordeaux and Erasmus for enabling the partnership between authors.  S.~N.~R acknowledges funding from the Programme Nationale de Planetologie (PNP) of the INSU (CNRS), and in the framework of the Investments for the Future programme IdEx, Universite de Bordeaux/RRI ORIGINS.

\printcredits

\bibliographystyle{cas-model2-names}

\bibliography{cas-refs}

\begin{thebibliography}{102}
\expandafter\ifx\csname natexlab\endcsname\relax\def\natexlab#1{#1}\fi
\providecommand{\url}[1]{\texttt{#1}}
\providecommand{\href}[2]{#2}
\providecommand{\path}[1]{#1}
\providecommand{\DOIprefix}{doi:}
\providecommand{\ArXivprefix}{arXiv:}
\providecommand{\URLprefix}{URL: }
\providecommand{\Pubmedprefix}{pmid:}
\providecommand{\doi}[1]{\href{http://dx.doi.org/#1}{\path{#1}}}
\providecommand{\Pubmed}[1]{\href{pmid:#1}{\path{#1}}}
\providecommand{\bibinfo}[2]{#2}
\ifx\xfnm\relax \def\xfnm[#1]{\unskip,\space#1}\fi
%Type = Article
\bibitem[{{Agnor} et~al.(1999){Agnor}, {Canup} and {Levison}}]{Agnor1999}
\bibinfo{author}{{Agnor}, C.B.}, \bibinfo{author}{{Canup}, R.M.},
  \bibinfo{author}{{Levison}, H.F.}, \bibinfo{year}{1999}.
\newblock \bibinfo{title}{{On the Character and Consequences of Large Impacts
  in the Late Stage of Terrestrial Planet Formation}}.
\newblock \bibinfo{journal}{Icarus} \bibinfo{volume}{142},
  \bibinfo{pages}{219--237}.
\newblock \DOIprefix\doi{10.1006/icar.1999.6201}.
%Type = Article
\bibitem[{{Armitage} et~al.(2016){Armitage}, {Eisner} and
  {Simon}}]{Armitage2016}
\bibinfo{author}{{Armitage}, P.J.}, \bibinfo{author}{{Eisner}, J.A.},
  \bibinfo{author}{{Simon}, J.B.}, \bibinfo{year}{2016}.
\newblock \bibinfo{title}{{Prompt Planetesimal Formation beyond the Snow
  Line}}.
\newblock \bibinfo{journal}{The Astrophysical Journal Letters}
  \bibinfo{volume}{828}, \bibinfo{pages}{L2}.
\newblock \DOIprefix\doi{10.3847/2041-8205/828/1/L2},
  \href{http://arxiv.org/abs/1608.03592}{\tt arXiv:1608.03592}.
%Type = Article
\bibitem[{{Avice} and {Marty}(2014)}]{Avice2014}
\bibinfo{author}{{Avice}, G.}, \bibinfo{author}{{Marty}, B.},
  \bibinfo{year}{2014}.
\newblock \bibinfo{title}{{The iodine-plutonium-xenon age of the Moon-Earth
  system revisited}}.
\newblock \bibinfo{journal}{Philosophical Transactions of the Royal Society of
  London Series A} \bibinfo{volume}{372}, \bibinfo{pages}{20130260--20130260}.
\newblock \DOIprefix\doi{10.1098/rsta.2013.0260},
  \href{http://arxiv.org/abs/1511.00952}{\tt arXiv:1511.00952}.
%Type = Article
\bibitem[{{Benz} et~al.(1986){Benz}, {Slattery} and {Cameron}}]{Benz1986}
\bibinfo{author}{{Benz}, W.}, \bibinfo{author}{{Slattery}, W.L.},
  \bibinfo{author}{{Cameron}, A.G.W.}, \bibinfo{year}{1986}.
\newblock \bibinfo{title}{{The origin of the moon and the single-impact
  hypothesis I}}.
\newblock \bibinfo{journal}{Icarus} \bibinfo{volume}{66},
  \bibinfo{pages}{515--535}.
\newblock \DOIprefix\doi{10.1016/0019-1035(86)90088-6}.
%Type = Article
\bibitem[{{Bermingham} et~al.(2025){Bermingham}, {Tornabene}, {Walker},
  {Godfrey}, {Meyer}, {Piccoli} and {Mojzsis}}]{Bermingham2025}
\bibinfo{author}{{Bermingham}, K.R.}, \bibinfo{author}{{Tornabene}, H.A.},
  \bibinfo{author}{{Walker}, R.J.}, \bibinfo{author}{{Godfrey}, L.V.},
  \bibinfo{author}{{Meyer}, B.S.}, \bibinfo{author}{{Piccoli}, P.},
  \bibinfo{author}{{Mojzsis}, S.J.}, \bibinfo{year}{2025}.
\newblock \bibinfo{title}{{The non-carbonaceous nature of Earth's late-stage
  accretion}}.
\newblock \bibinfo{journal}{Geochimica et Cosmochimica Acta}
  \bibinfo{volume}{392}, \bibinfo{pages}{38--51}.
\newblock \DOIprefix\doi{10.1016/j.gca.2024.11.005}.
%Type = Article
\bibitem[{{Birnstiel} et~al.(2016){Birnstiel}, {Fang} and
  {Johansen}}]{birnstiel16}
\bibinfo{author}{{Birnstiel}, T.}, \bibinfo{author}{{Fang}, M.},
  \bibinfo{author}{{Johansen}, A.}, \bibinfo{year}{2016}.
\newblock \bibinfo{title}{{Dust Evolution and the Formation of Planetesimals}}.
\newblock \bibinfo{journal}{Space Science Reviews} \bibinfo{volume}{205},
  \bibinfo{pages}{41--75}.
\newblock \DOIprefix\doi{10.1007/s11214-016-0256-1},
  \href{http://arxiv.org/abs/1604.02952}{\tt arXiv:1604.02952}.
%Type = Article
\bibitem[{{Bottke} et~al.(2010a){Bottke}, {Walker}, {Day}, {Nesvorny} and
  {Elkins-Tanton}}]{bottke10}
\bibinfo{author}{{Bottke}, W.F.}, \bibinfo{author}{{Walker}, R.J.},
  \bibinfo{author}{{Day}, J.M.D.}, \bibinfo{author}{{Nesvorny}, D.},
  \bibinfo{author}{{Elkins-Tanton}, L.}, \bibinfo{year}{2010}a.
\newblock \bibinfo{title}{{Stochastic Late Accretion to Earth, the Moon, and
  Mars}}.
\newblock \bibinfo{journal}{Science} \bibinfo{volume}{330},
  \bibinfo{pages}{1527}.
\newblock \DOIprefix\doi{10.1126/science.1196874}.
%Type = Article
\bibitem[{{Bottke} et~al.(2010b){Bottke}, {Walker}, {Day}, {Nesvorny} and
  {Elkins-Tanton}}]{Bottke2010}
\bibinfo{author}{{Bottke}, W.F.}, \bibinfo{author}{{Walker}, R.J.},
  \bibinfo{author}{{Day}, J.M.D.}, \bibinfo{author}{{Nesvorny}, D.},
  \bibinfo{author}{{Elkins-Tanton}, L.}, \bibinfo{year}{2010}b.
\newblock \bibinfo{title}{{Stochastic Late Accretion to Earth, the Moon, and
  Mars}}.
\newblock \bibinfo{journal}{Science} \bibinfo{volume}{330},
  \bibinfo{pages}{1527}.
\newblock \DOIprefix\doi{10.1126/science.1196874}.
%Type = Article
\bibitem[{{Brasser} and {Mojzsis}(2020)}]{Brasser2020}
\bibinfo{author}{{Brasser}, R.}, \bibinfo{author}{{Mojzsis}, S.J.},
  \bibinfo{year}{2020}.
\newblock \bibinfo{title}{{The partitioning of the inner and outer Solar System
  by a structured protoplanetary disk}}.
\newblock \bibinfo{journal}{Nature Astronomy} \bibinfo{volume}{4},
  \bibinfo{pages}{492--499}.
\newblock \DOIprefix\doi{10.1038/s41550-019-0978-6}.
%Type = Article
\bibitem[{Budde et~al.(2019)Budde, Burkhardt and Kleine}]{Budde2019}
\bibinfo{author}{Budde, G.}, \bibinfo{author}{Burkhardt, C.},
  \bibinfo{author}{Kleine, T.}, \bibinfo{year}{2019}.
\newblock \bibinfo{title}{Molybdenum isotopic evidence for the late accretion
  of outer solar system material to earth}.
\newblock \bibinfo{journal}{Nature Astronomy} \bibinfo{volume}{3},
  \bibinfo{pages}{736--741}.
\newblock \URLprefix \url{https://doi.org/10.1038/s41550-019-0779-y},
  \DOIprefix\doi{10.1038/s41550-019-0779-y}.
%Type = Article
\bibitem[{{Canup} et~al.(2021){Canup}, {Dauphas}, {Pahlevan}, {{\'C}uk},
  {Lock}, {Stewart}, {Salmon}, {Rufu}, {Nakajima} and
  {Magna}}]{canup2021origin}
\bibinfo{author}{{Canup}, Robin M.~{Righter}, K.}, \bibinfo{author}{{Dauphas},
  N.}, \bibinfo{author}{{Pahlevan}, K.}, \bibinfo{author}{{{\'C}uk}, M.},
  \bibinfo{author}{{Lock}, S.J.}, \bibinfo{author}{{Stewart}, S.T.},
  \bibinfo{author}{{Salmon}, J.}, \bibinfo{author}{{Rufu}, R.},
  \bibinfo{author}{{Nakajima}, M.}, \bibinfo{author}{{Magna}, T.},
  \bibinfo{year}{2021}.
\newblock \bibinfo{title}{{Origin of the Moon}}.
\newblock \bibinfo{journal}{arXiv e-prints}
  \href{http://arxiv.org/abs/2103.02045}{\tt arXiv:2103.02045}.
%Type = Article
\bibitem[{{Canup}(2012)}]{Canup2012}
\bibinfo{author}{{Canup}, R.M.}, \bibinfo{year}{2012}.
\newblock \bibinfo{title}{{Forming a Moon with an Earth-like Composition via a
  Giant Impact}}.
\newblock \bibinfo{journal}{Science} \bibinfo{volume}{338},
  \bibinfo{pages}{1052}.
\newblock \DOIprefix\doi{10.1126/science.1226073}.
%Type = Article
\bibitem[{{Canup} and {Asphaug}(2001a)}]{CanupAsphaug2001}
\bibinfo{author}{{Canup}, R.M.}, \bibinfo{author}{{Asphaug}, E.},
  \bibinfo{year}{2001}a.
\newblock \bibinfo{title}{{Origin of the Moon in a giant impact near the end of
  the Earth's formation}}.
\newblock \bibinfo{journal}{Nature} \bibinfo{volume}{412},
  \bibinfo{pages}{708--712}.
\newblock \DOIprefix\doi{10.1038/35089010}.
%Type = Article
\bibitem[{{Canup} and {Asphaug}(2001b)}]{Canup2001}
\bibinfo{author}{{Canup}, R.M.}, \bibinfo{author}{{Asphaug}, E.},
  \bibinfo{year}{2001}b.
\newblock \bibinfo{title}{{Origin of the Moon in a giant impact near the end of
  the Earth's formation}}.
\newblock \bibinfo{journal}{Nature} \bibinfo{volume}{412},
  \bibinfo{pages}{708--712}.
\newblock \DOIprefix\doi{10.1038/35089010}.
%Type = Article
\bibitem[{{Chambers}(1999)}]{Chambers1999}
\bibinfo{author}{{Chambers}, J.}, \bibinfo{year}{1999}.
\newblock \bibinfo{title}{{A hybrid symplectic integrator that permits close
  encounters between massive bodies}}.
\newblock \bibinfo{journal}{Monthly Notices of the Royal Astronomical Society}
  \bibinfo{volume}{304}, \bibinfo{pages}{793--799}.
\newblock \DOIprefix\doi{10.1046/j.1365-8711.1999.02379.x}.
%Type = Article
\bibitem[{Chambers and Wetherill(1998)}]{CHAMBERS1998304}
\bibinfo{author}{Chambers, J.}, \bibinfo{author}{Wetherill, G.},
  \bibinfo{year}{1998}.
\newblock \bibinfo{title}{Making the terrestrial planets: N-body integrations
  of planetary embryos in three dimensions}.
\newblock \bibinfo{journal}{Icarus} \bibinfo{volume}{136},
  \bibinfo{pages}{304--327}.
\newblock \URLprefix
  \url{https://www.sciencedirect.com/science/article/pii/S0019103598960079},
  \DOIprefix\doi{https://doi.org/10.1006/icar.1998.6007}.
%Type = Article
\bibitem[{{Chambers}(2001)}]{Chambers2001}
\bibinfo{author}{{Chambers}, J.E.}, \bibinfo{year}{2001}.
\newblock \bibinfo{title}{{Making More Terrestrial Planets}}.
\newblock \bibinfo{journal}{Icarus} \bibinfo{volume}{152},
  \bibinfo{pages}{205--224}.
\newblock \DOIprefix\doi{10.1006/icar.2001.6639}.
%Type = Article
\bibitem[{{Chambers} et~al.(1996){Chambers}, {Wetherill} and
  {Boss}}]{Chambers1996}
\bibinfo{author}{{Chambers}, J.E.}, \bibinfo{author}{{Wetherill}, G.W.},
  \bibinfo{author}{{Boss}, A.P.}, \bibinfo{year}{1996}.
\newblock \bibinfo{title}{{The Stability of Multi-Planet Systems}}.
\newblock \bibinfo{journal}{Icarus} \bibinfo{volume}{119},
  \bibinfo{pages}{261--268}.
\newblock \DOIprefix\doi{10.1006/icar.1996.0019}.
%Type = Article
\bibitem[{{Clement} et~al.(2018){Clement}, {Kaib}, {Raymond} and
  {Walsh}}]{Clement2018}
\bibinfo{author}{{Clement}, M.S.}, \bibinfo{author}{{Kaib}, N.A.},
  \bibinfo{author}{{Raymond}, S.N.}, \bibinfo{author}{{Walsh}, K.J.},
  \bibinfo{year}{2018}.
\newblock \bibinfo{title}{{Mars' growth stunted by an early giant planet
  instability}}.
\newblock \bibinfo{journal}{Icarus} \bibinfo{volume}{311},
  \bibinfo{pages}{340--356}.
\newblock \DOIprefix\doi{10.1016/j.icarus.2018.04.008},
  \href{http://arxiv.org/abs/1804.04233}{\tt arXiv:1804.04233}.
%Type = Article
\bibitem[{{{\'C}uk} and {Stewart}(2012)}]{Cuk2012}
\bibinfo{author}{{{\'C}uk}, M.}, \bibinfo{author}{{Stewart}, S.T.},
  \bibinfo{year}{2012}.
\newblock \bibinfo{title}{{Making the Moon from a Fast-Spinning Earth: A Giant
  Impact Followed by Resonant Despinning}}.
\newblock \bibinfo{journal}{Science} \bibinfo{volume}{338},
  \bibinfo{pages}{1047}.
\newblock \DOIprefix\doi{10.1126/science.1225542}.
%Type = Article
\bibitem[{{Dauphas}(2017)}]{dauphas17}
\bibinfo{author}{{Dauphas}, N.}, \bibinfo{year}{2017}.
\newblock \bibinfo{title}{{The isotopic nature of the Earth{\textquoteright}s
  accreting material through time}}.
\newblock \bibinfo{journal}{Nature} \bibinfo{volume}{541},
  \bibinfo{pages}{521--524}.
\newblock \DOIprefix\doi{10.1038/nature20830}.
%Type = Article
\bibitem[{{Dauphas} et~al.(2024){Dauphas}, {Hopp} and
  {Nesvorn{\'y}}}]{Dauphas2024}
\bibinfo{author}{{Dauphas}, N.}, \bibinfo{author}{{Hopp}, T.},
  \bibinfo{author}{{Nesvorn{\'y}}, D.}, \bibinfo{year}{2024}.
\newblock \bibinfo{title}{{Bayesian inference on the isotopic building blocks
  of Mars and Earth}}.
\newblock \bibinfo{journal}{Icarus} \bibinfo{volume}{408},
  \bibinfo{pages}{115805}.
\newblock \DOIprefix\doi{10.1016/j.icarus.2023.115805},
  \href{http://arxiv.org/abs/2309.15290}{\tt arXiv:2309.15290}.
%Type = Article
\bibitem[{{Day} et~al.(2007){Day}, {Pearson} and {Taylor}}]{day07}
\bibinfo{author}{{Day}, J.M.D.}, \bibinfo{author}{{Pearson}, D.G.},
  \bibinfo{author}{{Taylor}, L.A.}, \bibinfo{year}{2007}.
\newblock \bibinfo{title}{{Highly Siderophile Element Constraints on Accretion
  and Differentiation of the Earth-Moon System}}.
\newblock \bibinfo{journal}{Science} \bibinfo{volume}{315},
  \bibinfo{pages}{217}.
\newblock \DOIprefix\doi{10.1126/science.1133355}.
%Type = Article
\bibitem[{{DeSouza} et~al.(2021){DeSouza}, {Roig} and
  {Nesvorn{\'y}}}]{DeSouza2021}
\bibinfo{author}{{DeSouza}, S.R.}, \bibinfo{author}{{Roig}, F.},
  \bibinfo{author}{{Nesvorn{\'y}}, D.}, \bibinfo{year}{2021}.
\newblock \bibinfo{title}{{Can a jumping-Jupiter trigger the Moon's formation
  impact?}}
\newblock \bibinfo{journal}{Monthly Notices of the Royal Astronomical Society}
  \bibinfo{volume}{507}, \bibinfo{pages}{539--547}.
\newblock \DOIprefix\doi{10.1093/mnras/stab2188},
  \href{http://arxiv.org/abs/2107.04181}{\tt arXiv:2107.04181}.
%Type = Article
\bibitem[{{Dr{\k{a}}{\.z}kowska} and {Alibert}(2017)}]{Drazkowska2017}
\bibinfo{author}{{Dr{\k{a}}{\.z}kowska}, J.}, \bibinfo{author}{{Alibert}, Y.},
  \bibinfo{year}{2017}.
\newblock \bibinfo{title}{{Planetesimal formation starts at the snow line}}.
\newblock \bibinfo{journal}{Astronomy and Astrophysics} \bibinfo{volume}{608},
  \bibinfo{pages}{A92}.
\newblock \DOIprefix\doi{10.1051/0004-6361/201731491},
  \href{http://arxiv.org/abs/1710.00009}{\tt arXiv:1710.00009}.
%Type = Article
\bibitem[{{Dr{\k{a}}{\.z}kowska} et~al.(2016){Dr{\k{a}}{\.z}kowska}, {Alibert}
  and {Moore}}]{drazkowska16}
\bibinfo{author}{{Dr{\k{a}}{\.z}kowska}, J.}, \bibinfo{author}{{Alibert}, Y.},
  \bibinfo{author}{{Moore}, B.}, \bibinfo{year}{2016}.
\newblock \bibinfo{title}{{Close-in planetesimal formation by pile-up of
  drifting pebbles}}.
\newblock \bibinfo{journal}{Astronomy \& Astrophysics} \bibinfo{volume}{594},
  \bibinfo{pages}{A105}.
\newblock \DOIprefix\doi{10.1051/0004-6361/201628983},
  \href{http://arxiv.org/abs/1607.05734}{\tt arXiv:1607.05734}.
%Type = Article
\bibitem[{{Dr{\k{a}}{\.z}kowska} and {Dullemond}(2018)}]{drazkowska18}
\bibinfo{author}{{Dr{\k{a}}{\.z}kowska}, J.}, \bibinfo{author}{{Dullemond},
  C.P.}, \bibinfo{year}{2018}.
\newblock \bibinfo{title}{{Planetesimal formation during protoplanetary disk
  buildup}}.
\newblock \bibinfo{journal}{Astronomy \& Astrophysics} \bibinfo{volume}{614},
  \bibinfo{pages}{A62}.
\newblock \DOIprefix\doi{10.1051/0004-6361/201732221},
  \href{http://arxiv.org/abs/1803.00575}{\tt arXiv:1803.00575}.
%Type = Article
\bibitem[{{Edwards} et~al.(2024){Edwards}, {Keller}, {Newton} and
  {Stewart}}]{Edwards2024}
\bibinfo{author}{{Edwards}, G.H.}, \bibinfo{author}{{Keller}, C.B.},
  \bibinfo{author}{{Newton}, E.R.}, \bibinfo{author}{{Stewart}, C.W.},
  \bibinfo{year}{2024}.
\newblock \bibinfo{title}{{An early giant planet instability recorded in
  asteroidal meteorites}}.
\newblock \bibinfo{journal}{Nature Astronomy} \bibinfo{volume}{8},
  \bibinfo{pages}{1264--1276}.
\newblock \DOIprefix\doi{10.1038/s41550-024-02340-6},
  \href{http://arxiv.org/abs/2309.10906}{\tt arXiv:2309.10906}.
%Type = Article
\bibitem[{{Fischer} and {Ciesla}(2014)}]{Fischer2014}
\bibinfo{author}{{Fischer}, R.A.}, \bibinfo{author}{{Ciesla}, F.J.},
  \bibinfo{year}{2014}.
\newblock \bibinfo{title}{{Dynamics of the terrestrial planets from a large
  number of N-body simulations}}.
\newblock \bibinfo{journal}{Earth and Planetary Science Letters}
  \bibinfo{volume}{392}, \bibinfo{pages}{28--38}.
\newblock \DOIprefix\doi{10.1016/j.epsl.2014.02.011}.
%Type = Article
\bibitem[{Fischer and Nimmo(2018)}]{Fischer2018}
\bibinfo{author}{Fischer, R.A.}, \bibinfo{author}{Nimmo, F.},
  \bibinfo{year}{2018}.
\newblock \bibinfo{title}{Effects of core formation on the hf-w isotopic
  composition of the earth and dating of the moon-forming impact}.
\newblock \bibinfo{journal}{Earth and Planetary Science Letters}
  \bibinfo{volume}{499}, \bibinfo{pages}{257--265}.
\newblock \URLprefix
  \url{https://www.sciencedirect.com/science/article/pii/S0012821X18304369},
  \DOIprefix\doi{https://doi.org/10.1016/j.epsl.2018.07.030}.
%Type = Article
\bibitem[{{Gillmann} et~al.(2020){Gillmann}, {Golabek}, {Raymond},
  {Sch{\"o}nb{\"a}chler}, {Tackley}, {Dehant} and {Debaille}}]{Gillmann2020}
\bibinfo{author}{{Gillmann}, C.}, \bibinfo{author}{{Golabek}, G.J.},
  \bibinfo{author}{{Raymond}, S.N.}, \bibinfo{author}{{Sch{\"o}nb{\"a}chler},
  M.}, \bibinfo{author}{{Tackley}, P.J.}, \bibinfo{author}{{Dehant}, V.},
  \bibinfo{author}{{Debaille}, V.}, \bibinfo{year}{2020}.
\newblock \bibinfo{title}{{Dry late accretion inferred from Venus's coupled
  atmosphere and internal evolution}}.
\newblock \bibinfo{journal}{Nature Geoscience} \bibinfo{volume}{13},
  \bibinfo{pages}{265--269}.
\newblock \DOIprefix\doi{10.1038/s41561-020-0561-x},
  \href{http://arxiv.org/abs/2010.07132}{\tt arXiv:2010.07132}.
%Type = Article
\bibitem[{{Gomes} et~al.(2005){Gomes}, {Levison}, {Tsiganis} and
  {Morbidelli}}]{Gomes2005}
\bibinfo{author}{{Gomes}, R.}, \bibinfo{author}{{Levison}, H.F.},
  \bibinfo{author}{{Tsiganis}, K.}, \bibinfo{author}{{Morbidelli}, A.},
  \bibinfo{year}{2005}.
\newblock \bibinfo{title}{{Origin of the cataclysmic Late Heavy Bombardment
  period of the terrestrial planets}}.
\newblock \bibinfo{journal}{Nature} \bibinfo{volume}{435},
  \bibinfo{pages}{466--469}.
\newblock \DOIprefix\doi{10.1038/nature03676}.
%Type = Article
\bibitem[{Hansen(2009)}]{Hansen_2009}
\bibinfo{author}{Hansen, B.M.S.}, \bibinfo{year}{2009}.
\newblock \bibinfo{title}{Formation of the terrestrial planets from a narrow
  annulus}.
\newblock \bibinfo{journal}{The Astrophysical Journal} \bibinfo{volume}{703},
  \bibinfo{pages}{1131--1140}.
\newblock \URLprefix \url{http://dx.doi.org/10.1088/0004-637X/703/1/1131},
  \DOIprefix\doi{10.1088/0004-637x/703/1/1131}.
%Type = Article
\bibitem[{{Hansen}(2009)}]{Hansen09}
\bibinfo{author}{{Hansen}, B.M.S.}, \bibinfo{year}{2009}.
\newblock \bibinfo{title}{{Formation of the Terrestrial Planets from a Narrow
  Annulus}}.
\newblock \bibinfo{journal}{The Astrophysical Journal} \bibinfo{volume}{703},
  \bibinfo{pages}{1131--1140}.
\newblock \DOIprefix\doi{10.1088/0004-637X/703/1/1131},
  \href{http://arxiv.org/abs/0908.0743}{\tt arXiv:0908.0743}.
%Type = Article
\bibitem[{{Hunt} et~al.(2022){Hunt}, {Theis}, {Rehk{\"a}mper}, {Benedix},
  {Andreasen} and {Sch{\"o}nb{\"a}chler}}]{Hunt2022}
\bibinfo{author}{{Hunt}, A.C.}, \bibinfo{author}{{Theis}, K.J.},
  \bibinfo{author}{{Rehk{\"a}mper}, M.}, \bibinfo{author}{{Benedix}, G.K.},
  \bibinfo{author}{{Andreasen}, R.}, \bibinfo{author}{{Sch{\"o}nb{\"a}chler},
  M.}, \bibinfo{year}{2022}.
\newblock \bibinfo{title}{{The dissipation of the solar nebula constrained by
  impacts and core cooling in planetesimals}}.
\newblock \bibinfo{journal}{Nature Astronomy} \bibinfo{volume}{6},
  \bibinfo{pages}{812--818}.
\newblock \DOIprefix\doi{10.1038/s41550-022-01675-2},
  \href{http://arxiv.org/abs/2211.08306}{\tt arXiv:2211.08306}.
%Type = Article
\bibitem[{{Izidoro} et~al.(2022a){Izidoro}, {Dasgupta}, {Raymond}, {Deienno},
  {Bitsch} and {Isella}}]{izidoro22}
\bibinfo{author}{{Izidoro}, A.}, \bibinfo{author}{{Dasgupta}, R.},
  \bibinfo{author}{{Raymond}, S.N.}, \bibinfo{author}{{Deienno}, R.},
  \bibinfo{author}{{Bitsch}, B.}, \bibinfo{author}{{Isella}, A.},
  \bibinfo{year}{2022}a.
\newblock \bibinfo{title}{{Planetesimal rings as the cause of the Solar
  System's planetary architecture}}.
\newblock \bibinfo{journal}{Nature Astronomy} \bibinfo{volume}{6},
  \bibinfo{pages}{357--366}.
\newblock \DOIprefix\doi{10.1038/s41550-021-01557-z},
  \href{http://arxiv.org/abs/2112.15558}{\tt arXiv:2112.15558}.
%Type = Article
\bibitem[{{Izidoro} et~al.(2022b){Izidoro}, {Dasgupta}, {Raymond}, {Deienno},
  {Bitsch} and {Isella}}]{Izidoro2022}
\bibinfo{author}{{Izidoro}, A.}, \bibinfo{author}{{Dasgupta}, R.},
  \bibinfo{author}{{Raymond}, S.N.}, \bibinfo{author}{{Deienno}, R.},
  \bibinfo{author}{{Bitsch}, B.}, \bibinfo{author}{{Isella}, A.},
  \bibinfo{year}{2022}b.
\newblock \bibinfo{title}{{Planetesimal rings as the cause of the Solar
  System's planetary architecture}}.
\newblock \bibinfo{journal}{Nature Astronomy} \bibinfo{volume}{6},
  \bibinfo{pages}{357--366}.
\newblock \DOIprefix\doi{10.1038/s41550-021-01557-z},
  \href{http://arxiv.org/abs/2112.15558}{\tt arXiv:2112.15558}.
%Type = Article
\bibitem[{{Izidoro} et~al.(2024){Izidoro}, {Deienno}, {Raymond} and
  {Clement}}]{Izidoro2024}
\bibinfo{author}{{Izidoro}, A.}, \bibinfo{author}{{Deienno}, R.},
  \bibinfo{author}{{Raymond}, S.N.}, \bibinfo{author}{{Clement}, M.S.},
  \bibinfo{year}{2024}.
\newblock \bibinfo{title}{{Implantation of asteroids from the terrestrial
  planet region: The effect of the timing of the giant planet instability}}.
\newblock \bibinfo{journal}{arXiv e-prints} ,
  \bibinfo{pages}{arXiv:2404.10831}\DOIprefix\doi{10.48550/arXiv.2404.10831},
  \href{http://arxiv.org/abs/2404.10831}{\tt arXiv:2404.10831}.
%Type = Article
\bibitem[{{Izidoro} et~al.(2015){Izidoro}, {Raymond}, {Morbidelli} and
  {Winter}}]{Izidoro2015}
\bibinfo{author}{{Izidoro}, A.}, \bibinfo{author}{{Raymond}, S.N.},
  \bibinfo{author}{{Morbidelli}, A.}, \bibinfo{author}{{Winter}, O.C.},
  \bibinfo{year}{2015}.
\newblock \bibinfo{title}{{Terrestrial planet formation constrained by Mars and
  the structure of the asteroid belt}}.
\newblock \bibinfo{journal}{Monthly Notices of the Royal Astronomical Society}
  \bibinfo{volume}{453}, \bibinfo{pages}{3619--3634}.
\newblock \DOIprefix\doi{10.1093/mnras/stv1835},
  \href{http://arxiv.org/abs/1508.01365}{\tt arXiv:1508.01365}.
%Type = Article
\bibitem[{Jacobson et~al.(2014)Jacobson, Morbidelli, Raymond, O'Brien, Walsh
  and Rubie}]{Jacobson2014}
\bibinfo{author}{Jacobson, S.A.}, \bibinfo{author}{Morbidelli, A.},
  \bibinfo{author}{Raymond, S.N.}, \bibinfo{author}{O'Brien, D.P.},
  \bibinfo{author}{Walsh, K.J.}, \bibinfo{author}{Rubie, D.C.},
  \bibinfo{year}{2014}.
\newblock \bibinfo{title}{Highly siderophile elements in earth's mantle as a
  clock for the moon-forming impact}.
\newblock \bibinfo{journal}{Nature} \bibinfo{volume}{508},
  \bibinfo{pages}{84--87}.
\newblock \URLprefix \url{https://doi.org/10.1038/nature13172},
  \DOIprefix\doi{10.1038/nature13172}.
%Type = Inproceedings
\bibitem[{{Johansen} et~al.(2014){Johansen}, {Blum}, {Tanaka}, {Ormel},
  {Bizzarro} and {Rickman}}]{johansen14}
\bibinfo{author}{{Johansen}, A.}, \bibinfo{author}{{Blum}, J.},
  \bibinfo{author}{{Tanaka}, H.}, \bibinfo{author}{{Ormel}, C.},
  \bibinfo{author}{{Bizzarro}, M.}, \bibinfo{author}{{Rickman}, H.},
  \bibinfo{year}{2014}.
\newblock \bibinfo{title}{{The Multifaceted Planetesimal Formation Process}},
  in: \bibinfo{editor}{{Beuther}, H.}, \bibinfo{editor}{{Klessen}, R.S.},
  \bibinfo{editor}{{Dullemond}, C.P.}, \bibinfo{editor}{{Henning}, T.} (Eds.),
  \bibinfo{booktitle}{Protostars and Planets VI}, pp.
  \bibinfo{pages}{547--570}.
\newblock \DOIprefix\doi{10.2458/azu_uapress_9780816531240-ch024},
  \href{http://arxiv.org/abs/1402.1344}{\tt arXiv:1402.1344}.
%Type = Article
\bibitem[{Joiret et~al.(2024)Joiret, Raymond, Avice and Clement}]{JOIRET2024}
\bibinfo{author}{Joiret, S.}, \bibinfo{author}{Raymond, S.N.},
  \bibinfo{author}{Avice, G.}, \bibinfo{author}{Clement, M.S.},
  \bibinfo{year}{2024}.
\newblock \bibinfo{title}{Crash chronicles: Relative contribution from comets
  and carbonaceous asteroids to earth's volatile budget in the context of an
  early instability}.
\newblock \bibinfo{journal}{Icarus} \bibinfo{volume}{414},
  \bibinfo{pages}{116032}.
\newblock \URLprefix
  \url{https://www.sciencedirect.com/science/article/pii/S0019103524000915},
  \DOIprefix\doi{https://doi.org/10.1016/j.icarus.2024.116032}.
%Type = Article
\bibitem[{Kaib and Cowan(2015)}]{Kaib_2015}
\bibinfo{author}{Kaib, N.A.}, \bibinfo{author}{Cowan, N.B.},
  \bibinfo{year}{2015}.
\newblock \bibinfo{title}{The feeding zones of terrestrial planets and insights
  into moon formation}.
\newblock \bibinfo{journal}{Icarus} \bibinfo{volume}{252},
  \bibinfo{pages}{161--174}.
\newblock \URLprefix \url{http://dx.doi.org/10.1016/j.icarus.2015.01.013},
  \DOIprefix\doi{10.1016/j.icarus.2015.01.013}.
%Type = Article
\bibitem[{Kleine et~al.(2023)Kleine, Steller, Burkhardt and Nimmo}]{KLEINE2023}
\bibinfo{author}{Kleine, T.}, \bibinfo{author}{Steller, T.},
  \bibinfo{author}{Burkhardt, C.}, \bibinfo{author}{Nimmo, F.},
  \bibinfo{year}{2023}.
\newblock \bibinfo{title}{An inner solar system origin of volatile elements in
  mars}.
\newblock \bibinfo{journal}{Icarus} \bibinfo{volume}{397},
  \bibinfo{pages}{115519}.
\newblock \URLprefix
  \url{https://www.sciencedirect.com/science/article/pii/S0019103523000969},
  \DOIprefix\doi{https://doi.org/10.1016/j.icarus.2023.115519}.
%Type = Article
\bibitem[{{Kleine} et~al.(2009){Kleine}, {Touboul}, {Bourdon}, {Nimmo},
  {Mezger}, {Palme}, {Jacobsen}, {Yin} and {Halliday}}]{Kleine09}
\bibinfo{author}{{Kleine}, T.}, \bibinfo{author}{{Touboul}, M.},
  \bibinfo{author}{{Bourdon}, B.}, \bibinfo{author}{{Nimmo}, F.},
  \bibinfo{author}{{Mezger}, K.}, \bibinfo{author}{{Palme}, H.},
  \bibinfo{author}{{Jacobsen}, S.B.}, \bibinfo{author}{{Yin}, Q.Z.},
  \bibinfo{author}{{Halliday}, A.N.}, \bibinfo{year}{2009}.
\newblock \bibinfo{title}{{Hf-W chronology of the accretion and early evolution
  of asteroids and terrestrial planets}}.
\newblock \bibinfo{journal}{Geochimica et Cosmochimica Acta}
  \bibinfo{volume}{73}, \bibinfo{pages}{5150--5188}.
\newblock \DOIprefix\doi{10.1016/j.gca.2008.11.047}.
%Type = Article
\bibitem[{Kokubo and Ida(1998)}]{KokuboIda1998}
\bibinfo{author}{Kokubo, E.}, \bibinfo{author}{Ida, S.}, \bibinfo{year}{1998}.
\newblock \bibinfo{title}{Oligarchic growth of protoplanets}.
\newblock \bibinfo{journal}{Icarus} \bibinfo{volume}{131},
  \bibinfo{pages}{171--178}.
\newblock \URLprefix
  \url{https://www.sciencedirect.com/science/article/pii/S0019103597958401},
  \DOIprefix\doi{https://doi.org/10.1006/icar.1997.5840}.
%Type = Article
\bibitem[{{Kokubo} and {Ida}(2000)}]{kokubo00}
\bibinfo{author}{{Kokubo}, E.}, \bibinfo{author}{{Ida}, S.},
  \bibinfo{year}{2000}.
\newblock \bibinfo{title}{{Formation of Protoplanets from Planetesimals in the
  Solar Nebula}}.
\newblock \bibinfo{journal}{Icarus} \bibinfo{volume}{143},
  \bibinfo{pages}{15--27}.
\newblock \DOIprefix\doi{10.1006/icar.1999.6237}.
%Type = Article
\bibitem[{Kruijer et~al.(2017)Kruijer, Burkhardt, Budde and
  Kleine}]{KRUIJER2017}
\bibinfo{author}{Kruijer, T.S.}, \bibinfo{author}{Burkhardt, C.},
  \bibinfo{author}{Budde, G.}, \bibinfo{author}{Kleine, T.},
  \bibinfo{year}{2017}.
\newblock \bibinfo{title}{Age of jupiter inferred from the distinct genetics
  and formation times of meteorites}.
\newblock \bibinfo{journal}{Proceedings of the National Academy of Sciences}
  \bibinfo{volume}{114}, \bibinfo{pages}{6712--6716}.
\newblock \URLprefix
  \url{https://www.pnas.org/doi/abs/10.1073/pnas.1704461114},
  \DOIprefix\doi{https://doi.org/10.1073/pnas.1704461114},
  \href{http://arxiv.org/abs/https://www.pnas.org/doi/pdf/10.1073/pnas.1704461114}{\tt
  arXiv:https://www.pnas.org/doi/pdf/10.1073/pnas.1704461114}.
%Type = Article
\bibitem[{{Kruijer} et~al.(2020){Kruijer}, {Kleine} and {Borg}}]{Kruijer2020}
\bibinfo{author}{{Kruijer}, T.S.}, \bibinfo{author}{{Kleine}, T.},
  \bibinfo{author}{{Borg}, L.E.}, \bibinfo{year}{2020}.
\newblock \bibinfo{title}{{The great isotopic dichotomy of the early Solar
  System}}.
\newblock \bibinfo{journal}{Nature Astronomy} \bibinfo{volume}{4},
  \bibinfo{pages}{32--40}.
\newblock \DOIprefix\doi{10.1038/s41550-019-0959-9}.
%Type = Article
\bibitem[{{Lambrechts} and {Johansen}(2012)}]{Lambrechts2012}
\bibinfo{author}{{Lambrechts}, M.}, \bibinfo{author}{{Johansen}, A.},
  \bibinfo{year}{2012}.
\newblock \bibinfo{title}{{Rapid growth of gas-giant cores by pebble
  accretion}}.
\newblock \bibinfo{journal}{Astronomy \& Astrophysics} \bibinfo{volume}{544},
  \bibinfo{pages}{A32}.
\newblock \DOIprefix\doi{10.1051/0004-6361/201219127},
  \href{http://arxiv.org/abs/1205.3030}{\tt arXiv:1205.3030}.
%Type = Article
\bibitem[{{Laskar}(1997)}]{Lascar1997}
\bibinfo{author}{{Laskar}, J.}, \bibinfo{year}{1997}.
\newblock \bibinfo{title}{{Large scale chaos and the spacing of the inner
  planets.}}
\newblock \bibinfo{journal}{Astronomy and Astrophysics} \bibinfo{volume}{317},
  \bibinfo{pages}{L75--L78}.
%Type = Article
\bibitem[{{Leinhardt} and {Richardson}(2005)}]{leinhardt05}
\bibinfo{author}{{Leinhardt}, Z.M.}, \bibinfo{author}{{Richardson}, D.C.},
  \bibinfo{year}{2005}.
\newblock \bibinfo{title}{{Planetesimals to Protoplanets. I. Effect of
  Fragmentation on Terrestrial Planet Formation}}.
\newblock \bibinfo{journal}{The Astrophysical Journal} \bibinfo{volume}{625},
  \bibinfo{pages}{427--440}.
\newblock \DOIprefix\doi{10.1086/429402}.
%Type = Article
\bibitem[{{Levison} et~al.(2011){Levison}, {Morbidelli}, {Tsiganis},
  {Nesvorn{\'y}} and {Gomes}}]{Levison2011}
\bibinfo{author}{{Levison}, H.F.}, \bibinfo{author}{{Morbidelli}, A.},
  \bibinfo{author}{{Tsiganis}, K.}, \bibinfo{author}{{Nesvorn{\'y}}, D.},
  \bibinfo{author}{{Gomes}, R.}, \bibinfo{year}{2011}.
\newblock \bibinfo{title}{{Late Orbital Instabilities in the Outer Planets
  Induced by Interaction with a Self-gravitating Planetesimal Disk}}.
\newblock \bibinfo{journal}{The Astronomical Journal} \bibinfo{volume}{142},
  \bibinfo{pages}{152}.
\newblock \DOIprefix\doi{10.1088/0004-6256/142/5/152}.
%Type = Article
\bibitem[{{Lichtenberg} et~al.(2021){Lichtenberg}, {Dr{\k{a}}{\.z}kowska},
  {Sch{\"o}nb{\"a}chler}, {Golabek} and {Hands}}]{lichtenberg21}
\bibinfo{author}{{Lichtenberg}, T.}, \bibinfo{author}{{Dr{\k{a}}{\.z}kowska},
  J.}, \bibinfo{author}{{Sch{\"o}nb{\"a}chler}, M.},
  \bibinfo{author}{{Golabek}, G.J.}, \bibinfo{author}{{Hands}, T.O.},
  \bibinfo{year}{2021}.
\newblock \bibinfo{title}{{Bifurcation of planetary building blocks during
  Solar System formation}}.
\newblock \bibinfo{journal}{Science} \bibinfo{volume}{371},
  \bibinfo{pages}{365--370}.
\newblock \DOIprefix\doi{10.1126/science.abb3091},
  \href{http://arxiv.org/abs/2101.08571}{\tt arXiv:2101.08571}.
%Type = Article
\bibitem[{Liu et~al.(2022)Liu, Raymond and Jacobson}]{Liu2022}
\bibinfo{author}{Liu, B.}, \bibinfo{author}{Raymond, S.N.},
  \bibinfo{author}{Jacobson, S.A.}, \bibinfo{year}{2022}.
\newblock \bibinfo{title}{Early solar system instability triggered by dispersal
  of the gaseous disk}.
\newblock \bibinfo{journal}{Nature} \bibinfo{volume}{604},
  \bibinfo{pages}{643--646}.
\newblock \URLprefix \url{https://doi.org/10.1038/s41586-022-04535-1},
  \DOIprefix\doi{10.1038/s41586-022-04535-1}.
%Type = Article
\bibitem[{{Martins} et~al.(2023){Martins}, {Kuthning}, {Coles}, {Kreissig} and
  {Rehk{\"a}mper}}]{martins23}
\bibinfo{author}{{Martins}, R.}, \bibinfo{author}{{Kuthning}, S.},
  \bibinfo{author}{{Coles}, B.J.}, \bibinfo{author}{{Kreissig}, K.},
  \bibinfo{author}{{Rehk{\"a}mper}, M.}, \bibinfo{year}{2023}.
\newblock \bibinfo{title}{{Nucleosynthetic isotope anomalies of zinc in
  meteorites constrain the origin of Earth{\textquoteright}s volatiles}}.
\newblock \bibinfo{journal}{Science} \bibinfo{volume}{379},
  \bibinfo{pages}{369--372}.
\newblock \DOIprefix\doi{10.1126/science.abn1021}.
%Type = Article
\bibitem[{{Mojzsis} et~al.(2019a){Mojzsis}, {Brasser}, {Kelly}, {Abramov} and
  {Werner}}]{Mojzsis19}
\bibinfo{author}{{Mojzsis}, S.J.}, \bibinfo{author}{{Brasser}, R.},
  \bibinfo{author}{{Kelly}, N.M.}, \bibinfo{author}{{Abramov}, O.},
  \bibinfo{author}{{Werner}, S.C.}, \bibinfo{year}{2019}a.
\newblock \bibinfo{title}{{Onset of Giant Planet Migration before 4480 Million
  Years Ago}}.
\newblock \bibinfo{journal}{The Astrophysical Journal} \bibinfo{volume}{881},
  \bibinfo{pages}{44}.
\newblock \DOIprefix\doi{10.3847/1538-4357/ab2c03},
  \href{http://arxiv.org/abs/1903.08825}{\tt arXiv:1903.08825}.
%Type = Article
\bibitem[{{Mojzsis} et~al.(2019b){Mojzsis}, {Brasser}, {Kelly}, {Abramov} and
  {Werner}}]{Mojzsis2019}
\bibinfo{author}{{Mojzsis}, S.J.}, \bibinfo{author}{{Brasser}, R.},
  \bibinfo{author}{{Kelly}, N.M.}, \bibinfo{author}{{Abramov}, O.},
  \bibinfo{author}{{Werner}, S.C.}, \bibinfo{year}{2019}b.
\newblock \bibinfo{title}{{Onset of Giant Planet Migration before 4480 Million
  Years Ago}}.
\newblock \bibinfo{journal}{The Astrophysical Journal} \bibinfo{volume}{881},
  \bibinfo{pages}{44}.
\newblock \DOIprefix\doi{10.3847/1538-4357/ab2c03},
  \href{http://arxiv.org/abs/1903.08825}{\tt arXiv:1903.08825}.
%Type = Article
\bibitem[{{Morbidelli} et~al.(2022){Morbidelli}, {Bailli{\'e}}, {Batygin},
  {Charnoz}, {Guillot}, {Rubie} and {Kleine}}]{Morbidelli2022}
\bibinfo{author}{{Morbidelli}, A.}, \bibinfo{author}{{Bailli{\'e}}, K.},
  \bibinfo{author}{{Batygin}, K.}, \bibinfo{author}{{Charnoz}, S.},
  \bibinfo{author}{{Guillot}, T.}, \bibinfo{author}{{Rubie}, D.C.},
  \bibinfo{author}{{Kleine}, T.}, \bibinfo{year}{2022}.
\newblock \bibinfo{title}{{Contemporary formation of early Solar System
  planetesimals at two distinct radial locations}}.
\newblock \bibinfo{journal}{Nature Astronomy} \bibinfo{volume}{6},
  \bibinfo{pages}{72--79}.
\newblock \DOIprefix\doi{10.1038/s41550-021-01517-7},
  \href{http://arxiv.org/abs/2112.15413}{\tt arXiv:2112.15413}.
%Type = Article
\bibitem[{{Morbidelli} et~al.(2012){Morbidelli}, {Lunine}, {O'Brien}, {Raymond}
  and {Walsh}}]{Morbidelli2012}
\bibinfo{author}{{Morbidelli}, A.}, \bibinfo{author}{{Lunine}, J.I.},
  \bibinfo{author}{{O'Brien}, D.P.}, \bibinfo{author}{{Raymond}, S.N.},
  \bibinfo{author}{{Walsh}, K.J.}, \bibinfo{year}{2012}.
\newblock \bibinfo{title}{{Building Terrestrial Planets}}.
\newblock \bibinfo{journal}{Annual Review of Earth and Planetary Sciences}
  \bibinfo{volume}{40}, \bibinfo{pages}{251--275}.
\newblock \DOIprefix\doi{10.1146/annurev-earth-042711-105319},
  \href{http://arxiv.org/abs/1208.4694}{\tt arXiv:1208.4694}.
%Type = Article
\bibitem[{{Morbidelli} et~al.(2018){Morbidelli}, {Nesvorny}, {Laurenz},
  {Marchi}, {Rubie}, {Elkins-Tanton}, {Wieczorek} and {Jacobson}}]{morby18}
\bibinfo{author}{{Morbidelli}, A.}, \bibinfo{author}{{Nesvorny}, D.},
  \bibinfo{author}{{Laurenz}, V.}, \bibinfo{author}{{Marchi}, S.},
  \bibinfo{author}{{Rubie}, D.C.}, \bibinfo{author}{{Elkins-Tanton}, L.},
  \bibinfo{author}{{Wieczorek}, M.}, \bibinfo{author}{{Jacobson}, S.},
  \bibinfo{year}{2018}.
\newblock \bibinfo{title}{{The timeline of the lunar bombardment: Revisited}}.
\newblock \bibinfo{journal}{Icarus} \bibinfo{volume}{305},
  \bibinfo{pages}{262--276}.
\newblock \DOIprefix\doi{10.1016/j.icarus.2017.12.046},
  \href{http://arxiv.org/abs/1801.03756}{\tt arXiv:1801.03756}.
%Type = Article
\bibitem[{{Morbidelli} et~al.(2007){Morbidelli}, {Tsiganis}, {Crida}, {Levison}
  and {Gomes}}]{Morbidelli2007}
\bibinfo{author}{{Morbidelli}, A.}, \bibinfo{author}{{Tsiganis}, K.},
  \bibinfo{author}{{Crida}, A.}, \bibinfo{author}{{Levison}, H.F.},
  \bibinfo{author}{{Gomes}, R.}, \bibinfo{year}{2007}.
\newblock \bibinfo{title}{{Dynamics of the Giant Planets of the Solar System in
  the Gaseous Protoplanetary Disk and Their Relationship to the Current Orbital
  Architecture}}.
\newblock \bibinfo{journal}{The Astronomical Journal} \bibinfo{volume}{134},
  \bibinfo{pages}{1790--1798}.
\newblock \DOIprefix\doi{10.1086/521705},
  \href{http://arxiv.org/abs/0706.1713}{\tt arXiv:0706.1713}.
%Type = Article
\bibitem[{{Morbidelli} and {Wood}(2015a)}]{Morbidelli2015}
\bibinfo{author}{{Morbidelli}, A.}, \bibinfo{author}{{Wood}, B.J.},
  \bibinfo{year}{2015}a.
\newblock \bibinfo{title}{{Late Accretion and the Late Veneer}}.
\newblock \bibinfo{journal}{Geophysical Monograph Series}
  \bibinfo{volume}{212}, \bibinfo{pages}{71--82}.
\newblock \DOIprefix\doi{10.1002/9781118860359.ch4},
  \href{http://arxiv.org/abs/1411.4563}{\tt arXiv:1411.4563}.
%Type = Article
\bibitem[{{Morbidelli} and {Wood}(2015b)}]{MorbyWood2015}
\bibinfo{author}{{Morbidelli}, A.}, \bibinfo{author}{{Wood}, B.J.},
  \bibinfo{year}{2015}b.
\newblock \bibinfo{title}{{Late Accretion and the Late Veneer}}.
\newblock \bibinfo{journal}{Geophysical Monograph Series}
  \bibinfo{volume}{212}, \bibinfo{pages}{71--82}.
\newblock \DOIprefix\doi{10.1002/9781118860359.ch4},
  \href{http://arxiv.org/abs/1411.4563}{\tt arXiv:1411.4563}.
%Type = Article
\bibitem[{{Nesvorn{\'y}} et~al.(2024){Nesvorn{\'y}}, {Dauphas},
  {Vokrouhlick{\'y}}, {Deienno} and {Hopp}}]{Nesvorny2024}
\bibinfo{author}{{Nesvorn{\'y}}, D.}, \bibinfo{author}{{Dauphas}, N.},
  \bibinfo{author}{{Vokrouhlick{\'y}}, D.}, \bibinfo{author}{{Deienno}, R.},
  \bibinfo{author}{{Hopp}, T.}, \bibinfo{year}{2024}.
\newblock \bibinfo{title}{{Isotopic trichotomy of main belt asteroids from
  implantation of outer solar system planetesimals}}.
\newblock \bibinfo{journal}{Earth and Planetary Science Letters}
  \bibinfo{volume}{626}, \bibinfo{pages}{118521}.
\newblock \DOIprefix\doi{10.1016/j.epsl.2023.118521},
  \href{http://arxiv.org/abs/2311.16053}{\tt arXiv:2311.16053}.
%Type = Article
\bibitem[{{Nesvorn{\'y}} et~al.(2018){Nesvorn{\'y}}, {Vokrouhlick{\'y}},
  {Bottke} and {Levison}}]{Nesvorny18b}
\bibinfo{author}{{Nesvorn{\'y}}, D.}, \bibinfo{author}{{Vokrouhlick{\'y}}, D.},
  \bibinfo{author}{{Bottke}, W.F.}, \bibinfo{author}{{Levison}, H.F.},
  \bibinfo{year}{2018}.
\newblock \bibinfo{title}{{Evidence for very early migration of the Solar
  System planets from the Patroclus-Menoetius binary Jupiter Trojan}}.
\newblock \bibinfo{journal}{Nature Astronomy} \bibinfo{volume}{2},
  \bibinfo{pages}{878--882}.
\newblock \DOIprefix\doi{10.1038/s41550-018-0564-3},
  \href{http://arxiv.org/abs/1809.04007}{\tt arXiv:1809.04007}.
%Type = Article
\bibitem[{Nesvorný(2018)}]{Nesvorny2018}
\bibinfo{author}{Nesvorný, D.}, \bibinfo{year}{2018}.
\newblock \bibinfo{title}{Dynamical evolution of the early solar system}.
\newblock \bibinfo{journal}{Annual Review of Astronomy and Astrophysics}
  \bibinfo{volume}{56}, \bibinfo{pages}{137--174}.
\newblock \URLprefix
  \url{http://dx.doi.org/10.1146/annurev-astro-081817-052028},
  \DOIprefix\doi{10.1146/annurev-astro-081817-052028}.
%Type = Article
\bibitem[{Nesvorný et~al.(2021)Nesvorný, Roig and Deienno}]{Nesvorny2021}
\bibinfo{author}{Nesvorný, D.}, \bibinfo{author}{Roig, F.V.},
  \bibinfo{author}{Deienno, R.}, \bibinfo{year}{2021}.
\newblock \bibinfo{title}{The role of early giant-planet instability in
  terrestrial planet formation}.
\newblock \bibinfo{journal}{The Astronomical Journal} \bibinfo{volume}{161},
  \bibinfo{pages}{50}.
\newblock \URLprefix \url{https://dx.doi.org/10.3847/1538-3881/abc8ef},
  \DOIprefix\doi{10.3847/1538-3881/abc8ef}.
%Type = Article
\bibitem[{{Nimmo} et~al.(2024a){Nimmo}, {Kleine} and {Morbidelli}}]{Nimmo2024b}
\bibinfo{author}{{Nimmo}, F.}, \bibinfo{author}{{Kleine}, T.},
  \bibinfo{author}{{Morbidelli}, A.}, \bibinfo{year}{2024}a.
\newblock \bibinfo{title}{{Tidally driven remelting around 4.35 billion years
  ago indicates the Moon is old}}.
\newblock \bibinfo{journal}{Nature} \bibinfo{volume}{636},
  \bibinfo{pages}{598--602}.
\newblock \DOIprefix\doi{10.1038/s41586-024-08231-0}.
%Type = Article
\bibitem[{{Nimmo} et~al.(2024b){Nimmo}, {Kleine}, {Morbidelli} and
  {Nesvorny}}]{Nimmo2024}
\bibinfo{author}{{Nimmo}, F.}, \bibinfo{author}{{Kleine}, T.},
  \bibinfo{author}{{Morbidelli}, A.}, \bibinfo{author}{{Nesvorny}, D.},
  \bibinfo{year}{2024}b.
\newblock \bibinfo{title}{{Mechanisms and timing of carbonaceous chondrite
  delivery to the Earth}}.
\newblock \bibinfo{journal}{Earth and Planetary Science Letters}
  \bibinfo{volume}{648}, \bibinfo{pages}{119112}.
\newblock \DOIprefix\doi{10.1016/j.epsl.2024.119112},
  \href{http://arxiv.org/abs/2411.04889}{\tt arXiv:2411.04889}.
%Type = Article
\bibitem[{{O'Brien} et~al.(2006){O'Brien}, {Morbidelli} and
  {Levison}}]{OBRIEN2006}
\bibinfo{author}{{O'Brien}, D.P.}, \bibinfo{author}{{Morbidelli}, A.},
  \bibinfo{author}{{Levison}, H.F.}, \bibinfo{year}{2006}.
\newblock \bibinfo{title}{{Terrestrial planet formation with strong dynamical
  friction}}.
\newblock \bibinfo{journal}{Icarus} \bibinfo{volume}{184},
  \bibinfo{pages}{39--58}.
\newblock \DOIprefix\doi{10.1016/j.icarus.2006.04.005}.
%Type = Article
\bibitem[{O'Brien et~al.(2014)O'Brien, Walsh, Morbidelli, Raymond and
  Mandell}]{OBRIEN2014}
\bibinfo{author}{O'Brien, D.P.}, \bibinfo{author}{Walsh, K.J.},
  \bibinfo{author}{Morbidelli, A.}, \bibinfo{author}{Raymond, S.N.},
  \bibinfo{author}{Mandell, A.M.}, \bibinfo{year}{2014}.
\newblock \bibinfo{title}{Water delivery and giant impacts in the "grand tack"
  scenario}.
\newblock \bibinfo{journal}{Icarus} \bibinfo{volume}{239},
  \bibinfo{pages}{74--84}.
\newblock \URLprefix
  \url{https://www.sciencedirect.com/science/article/pii/S0019103514002620},
  \DOIprefix\doi{https://doi.org/10.1016/j.icarus.2014.05.009}.
%Type = Article
\bibitem[{{Ormel} and {Klahr}(2010)}]{Ormel2010}
\bibinfo{author}{{Ormel}, C.W.}, \bibinfo{author}{{Klahr}, H.H.},
  \bibinfo{year}{2010}.
\newblock \bibinfo{title}{{The effect of gas drag on the growth of
  protoplanets. Analytical expressions for the accretion of small bodies in
  laminar disks}}.
\newblock \bibinfo{journal}{Astronomy and Astrophysics} \bibinfo{volume}{520},
  \bibinfo{pages}{A43}.
\newblock \DOIprefix\doi{10.1051/0004-6361/201014903},
  \href{http://arxiv.org/abs/1007.0916}{\tt arXiv:1007.0916}.
%Type = Article
\bibitem[{{Pierens} and {Nelson}(2008)}]{Pierens2008}
\bibinfo{author}{{Pierens}, A.}, \bibinfo{author}{{Nelson}, R.P.},
  \bibinfo{year}{2008}.
\newblock \bibinfo{title}{{Constraints on resonant-trapping for two planets
  embedded in a protoplanetary disc}}.
\newblock \bibinfo{journal}{Astronomy and Astrophysics} \bibinfo{volume}{482},
  \bibinfo{pages}{333--340}.
\newblock \DOIprefix\doi{10.1051/0004-6361:20079062},
  \href{http://arxiv.org/abs/0802.2033}{\tt arXiv:0802.2033}.
%Type = Article
\bibitem[{{Pirani} et~al.(2019){Pirani}, {Johansen}, {Bitsch}, {Mustill} and
  {Turrini}}]{Pirani2019}
\bibinfo{author}{{Pirani}, S.}, \bibinfo{author}{{Johansen}, A.},
  \bibinfo{author}{{Bitsch}, B.}, \bibinfo{author}{{Mustill}, A.J.},
  \bibinfo{author}{{Turrini}, D.}, \bibinfo{year}{2019}.
\newblock \bibinfo{title}{{Consequences of planetary migration on the minor
  bodies of the early solar system}}.
\newblock \bibinfo{journal}{Astronomy and Astrophysics} \bibinfo{volume}{623},
  \bibinfo{pages}{A169}.
\newblock \DOIprefix\doi{10.1051/0004-6361/201833713},
  \href{http://arxiv.org/abs/1902.04591}{\tt arXiv:1902.04591}.
%Type = Article
\bibitem[{{Rauch} and {Holman}(1999)}]{Rauch_Holman99}
\bibinfo{author}{{Rauch}, K.P.}, \bibinfo{author}{{Holman}, M.},
  \bibinfo{year}{1999}.
\newblock \bibinfo{title}{{Dynamical Chaos in the Wisdom-Holman Integrator:
  Origins and Solutions}}.
\newblock \bibinfo{journal}{The Astronomical Journal} \bibinfo{volume}{117},
  \bibinfo{pages}{1087--1102}.
\newblock \DOIprefix\doi{10.1086/300720},
  \href{http://arxiv.org/abs/astro-ph/9803340}{\tt arXiv:astro-ph/9803340}.
%Type = Article
\bibitem[{Raymond and Izidoro(2017)}]{Raymond2017SciAdv}
\bibinfo{author}{Raymond, S.N.}, \bibinfo{author}{Izidoro, A.},
  \bibinfo{year}{2017}.
\newblock \bibinfo{title}{The empty primordial asteroid belt}.
\newblock \bibinfo{journal}{Science Advances} \bibinfo{volume}{3},
  \bibinfo{pages}{e1701138}.
\newblock \URLprefix
  \url{https://www.science.org/doi/abs/10.1126/sciadv.1701138},
  \DOIprefix\doi{10.1126/sciadv.1701138},
  \href{http://arxiv.org/abs/https://www.science.org/doi/pdf/10.1126/sciadv.1701138}{\tt
  arXiv:https://www.science.org/doi/pdf/10.1126/sciadv.1701138}.
%Type = Article
\bibitem[{{Raymond} and {Izidoro}(2017a)}]{RaymondIzidoro2017}
\bibinfo{author}{{Raymond}, S.N.}, \bibinfo{author}{{Izidoro}, A.},
  \bibinfo{year}{2017}a.
\newblock \bibinfo{title}{{Origin of water in the inner Solar System:
  Planetesimals scattered inward during Jupiter and Saturn's rapid gas
  accretion}}.
\newblock \bibinfo{journal}{Icarus} \bibinfo{volume}{297},
  \bibinfo{pages}{134--148}.
\newblock \DOIprefix\doi{10.1016/j.icarus.2017.06.030},
  \href{http://arxiv.org/abs/1707.01234}{\tt arXiv:1707.01234}.
%Type = Article
\bibitem[{{Raymond} and {Izidoro}(2017b)}]{RaymondIzidoro2017b}
\bibinfo{author}{{Raymond}, S.N.}, \bibinfo{author}{{Izidoro}, A.},
  \bibinfo{year}{2017}b.
\newblock \bibinfo{title}{{The empty primordial asteroid belt}}.
\newblock \bibinfo{journal}{Science Advances} \bibinfo{volume}{3},
  \bibinfo{pages}{e1701138}.
\newblock \DOIprefix\doi{10.1126/sciadv.1701138},
  \href{http://arxiv.org/abs/1709.04242}{\tt arXiv:1709.04242}.
%Type = Inbook
\bibitem[{Raymond et~al.(2014)Raymond, Kokubo, Morbidelli, Morishima and
  Walsh}]{Raymond2014}
\bibinfo{author}{Raymond, S.N.}, \bibinfo{author}{Kokubo, E.},
  \bibinfo{author}{Morbidelli, A.}, \bibinfo{author}{Morishima, R.},
  \bibinfo{author}{Walsh, K.J.}, \bibinfo{year}{2014}.
\newblock \bibinfo{title}{Terrestrial Planet Formation at Home and Abroad}.
  \bibinfo{publisher}{University of Arizona Press}.
\newblock \URLprefix
  \url{http://dx.doi.org/10.2458/azu_uapress_9780816531240-ch026},
  \DOIprefix\doi{10.2458/azu_uapress_9780816531240-ch026}.
%Type = Inproceedings
\bibitem[{{Raymond} and {Morbidelli}(2022)}]{raymond22}
\bibinfo{author}{{Raymond}, S.N.}, \bibinfo{author}{{Morbidelli}, A.},
  \bibinfo{year}{2022}.
\newblock \bibinfo{title}{{Planet Formation: Key Mechanisms and Global
  Models}}, in: \bibinfo{editor}{{Biazzo}, K.}, \bibinfo{editor}{{Bozza}, V.},
  \bibinfo{editor}{{Mancini}, L.}, \bibinfo{editor}{{Sozzetti}, A.} (Eds.),
  \bibinfo{booktitle}{Demographics of Exoplanetary Systems, Lecture Notes of
  the 3rd Advanced School on Exoplanetary Science}, pp. \bibinfo{pages}{3--82}.
\newblock \DOIprefix\doi{10.1007/978-3-030-88124-5_1},
  \href{http://arxiv.org/abs/2002.05756}{\tt arXiv:2002.05756}.
%Type = Article
\bibitem[{{Raymond} et~al.(2009){Raymond}, {O'Brien}, {Morbidelli} and
  {Kaib}}]{Raymond2009}
\bibinfo{author}{{Raymond}, S.N.}, \bibinfo{author}{{O'Brien}, D.P.},
  \bibinfo{author}{{Morbidelli}, A.}, \bibinfo{author}{{Kaib}, N.A.},
  \bibinfo{year}{2009}.
\newblock \bibinfo{title}{{Building the terrestrial planets: Constrained
  accretion in the inner Solar System}}.
\newblock \bibinfo{journal}{Icarus} \bibinfo{volume}{203},
  \bibinfo{pages}{644--662}.
\newblock \DOIprefix\doi{10.1016/j.icarus.2009.05.016},
  \href{http://arxiv.org/abs/0905.3750}{\tt arXiv:0905.3750}.
%Type = Article
\bibitem[{Raymond et~al.(2005)Raymond, Quinn and Lunine}]{Raymond_2005}
\bibinfo{author}{Raymond, S.N.}, \bibinfo{author}{Quinn, T.},
  \bibinfo{author}{Lunine, J.I.}, \bibinfo{year}{2005}.
\newblock \bibinfo{title}{Terrestrial planet formation in disks with varying
  surface density profiles}.
\newblock \bibinfo{journal}{The Astrophysical Journal} \bibinfo{volume}{632},
  \bibinfo{pages}{670--676}.
\newblock \URLprefix \url{http://dx.doi.org/10.1086/433179},
  \DOIprefix\doi{10.1086/433179}.
%Type = Article
\bibitem[{{Raymond} et~al.(2006){Raymond}, {Quinn} and {Lunine}}]{Raymond2006}
\bibinfo{author}{{Raymond}, S.N.}, \bibinfo{author}{{Quinn}, T.},
  \bibinfo{author}{{Lunine}, J.I.}, \bibinfo{year}{2006}.
\newblock \bibinfo{title}{{High-resolution simulations of the final assembly of
  Earth-like planets I. Terrestrial accretion and dynamics}}.
\newblock \bibinfo{journal}{Icarus} \bibinfo{volume}{183},
  \bibinfo{pages}{265--282}.
\newblock \DOIprefix\doi{10.1016/j.icarus.2006.03.011},
  \href{http://arxiv.org/abs/astro-ph/0510284}{\tt arXiv:astro-ph/0510284}.
%Type = Article
\bibitem[{{Raymond} et~al.(2013){Raymond}, {Schlichting}, {Hersant} and
  {Selsis}}]{raymond13}
\bibinfo{author}{{Raymond}, S.N.}, \bibinfo{author}{{Schlichting}, H.E.},
  \bibinfo{author}{{Hersant}, F.}, \bibinfo{author}{{Selsis}, F.},
  \bibinfo{year}{2013}.
\newblock \bibinfo{title}{{Dynamical and collisional constraints on a
  stochastic late veneer on the terrestrial planets}}.
\newblock \bibinfo{journal}{Icarus} \bibinfo{volume}{226},
  \bibinfo{pages}{671--681}.
\newblock \DOIprefix\doi{10.1016/j.icarus.2013.06.019},
  \href{http://arxiv.org/abs/1306.4325}{\tt arXiv:1306.4325}.
%Type = Article
\bibitem[{{Rubie} et~al.(2015){Rubie}, {Jacobson}, {Morbidelli}, {O'Brien},
  {Young}, {de Vries}, {Nimmo}, {Palme} and {Frost}}]{Rubie2015}
\bibinfo{author}{{Rubie}, D.C.}, \bibinfo{author}{{Jacobson}, S.A.},
  \bibinfo{author}{{Morbidelli}, A.}, \bibinfo{author}{{O'Brien}, D.P.},
  \bibinfo{author}{{Young}, E.D.}, \bibinfo{author}{{de Vries}, J.},
  \bibinfo{author}{{Nimmo}, F.}, \bibinfo{author}{{Palme}, H.},
  \bibinfo{author}{{Frost}, D.J.}, \bibinfo{year}{2015}.
\newblock \bibinfo{title}{{Accretion and differentiation of the terrestrial
  planets with implications for the compositions of early-formed Solar System
  bodies and accretion of water}}.
\newblock \bibinfo{journal}{Icarus} \bibinfo{volume}{248},
  \bibinfo{pages}{89--108}.
\newblock \DOIprefix\doi{10.1016/j.icarus.2014.10.015},
  \href{http://arxiv.org/abs/1410.3509}{\tt arXiv:1410.3509}.
%Type = Article
\bibitem[{{Savage} et~al.(2022){Savage}, {Moynier} and {Boyet}}]{savage2022}
\bibinfo{author}{{Savage}, P.S.}, \bibinfo{author}{{Moynier}, F.},
  \bibinfo{author}{{Boyet}, M.}, \bibinfo{year}{2022}.
\newblock \bibinfo{title}{{Zinc isotope anomalies in primitive meteorites
  identify the outer solar system as an important source of Earth's volatile
  inventory}}.
\newblock \bibinfo{journal}{Icarus} \bibinfo{volume}{386},
  \bibinfo{pages}{115172}.
\newblock \DOIprefix\doi{10.1016/j.icarus.2022.115172}.
%Type = Article
\bibitem[{de~Sousa~Ribeiro et~al.(2020)de~Sousa~Ribeiro, Morbidelli, Raymond,
  Izidoro, Gomes and {Vieira Neto}}]{RibeirodeSousa2020}
\bibinfo{author}{de~Sousa~Ribeiro, R.}, \bibinfo{author}{Morbidelli, A.},
  \bibinfo{author}{Raymond, S.N.}, \bibinfo{author}{Izidoro, A.},
  \bibinfo{author}{Gomes, R.}, \bibinfo{author}{{Vieira Neto}, E.},
  \bibinfo{year}{2020}.
\newblock \bibinfo{title}{Dynamical evidence for an early giant planet
  instability}.
\newblock \bibinfo{journal}{Icarus} \bibinfo{volume}{339},
  \bibinfo{pages}{113605}.
\newblock \URLprefix
  \url{https://www.sciencedirect.com/science/article/pii/S0019103519301332},
  \DOIprefix\doi{https://doi.org/10.1016/j.icarus.2019.113605}.
%Type = Article
\bibitem[{{Steller} et~al.(2022){Steller}, {Burkhardt}, {Yang} and
  {Kleine}}]{steller2022}
\bibinfo{author}{{Steller}, T.}, \bibinfo{author}{{Burkhardt}, C.},
  \bibinfo{author}{{Yang}, C.}, \bibinfo{author}{{Kleine}, T.},
  \bibinfo{year}{2022}.
\newblock \bibinfo{title}{{Nucleosynthetic zinc isotope anomalies reveal a dual
  origin of terrestrial volatiles}}.
\newblock \bibinfo{journal}{Icarus} \bibinfo{volume}{386},
  \bibinfo{pages}{115171}.
\newblock \DOIprefix\doi{10.1016/j.icarus.2022.115171}.
%Type = Article
\bibitem[{{Thiemens} et~al.(2019){Thiemens}, {Sprung}, {Fonseca}, {Leitzke} and
  {M{\"u}nker}}]{Thiemens2019}
\bibinfo{author}{{Thiemens}, M.M.}, \bibinfo{author}{{Sprung}, P.},
  \bibinfo{author}{{Fonseca}, R.O.C.}, \bibinfo{author}{{Leitzke}, F.P.},
  \bibinfo{author}{{M{\"u}nker}, C.}, \bibinfo{year}{2019}.
\newblock \bibinfo{title}{{Early Moon formation inferred from
  hafnium{\textendash}tungsten systematics}}.
\newblock \bibinfo{journal}{Nature Geoscience} \bibinfo{volume}{12},
  \bibinfo{pages}{696--700}.
\newblock \DOIprefix\doi{10.1038/s41561-019-0398-3}.
%Type = Article
\bibitem[{{Thommes} et~al.(2003){Thommes}, {Duncan} and
  {Levison}}]{Thommes2003}
\bibinfo{author}{{Thommes}, E.W.}, \bibinfo{author}{{Duncan}, M.J.},
  \bibinfo{author}{{Levison}, H.F.}, \bibinfo{year}{2003}.
\newblock \bibinfo{title}{{Oligarchic growth of giant planets}}.
\newblock \bibinfo{journal}{Icarus} \bibinfo{volume}{161},
  \bibinfo{pages}{431--455}.
\newblock \DOIprefix\doi{10.1016/S0019-1035(02)00043-X},
  \href{http://arxiv.org/abs/astro-ph/0303269}{\tt arXiv:astro-ph/0303269}.
%Type = Article
\bibitem[{Tsiganis et~al.(2005)Tsiganis, Gomes, Morbidelli and
  Levison}]{Tsiganis2005}
\bibinfo{author}{Tsiganis, K.}, \bibinfo{author}{Gomes, R.},
  \bibinfo{author}{Morbidelli, A.}, \bibinfo{author}{Levison, H.F.},
  \bibinfo{year}{2005}.
\newblock \bibinfo{title}{Origin of the orbital architecture of the giant
  planets of the solar system}.
\newblock \bibinfo{journal}{Nature} \bibinfo{volume}{435},
  \bibinfo{pages}{459--461}.
%Type = Article
\bibitem[{{Walker}(2009)}]{walker09}
\bibinfo{author}{{Walker}, R.J.}, \bibinfo{year}{2009}.
\newblock \bibinfo{title}{{Highly siderophile elements in the Earth, Moon and
  Mars: Update and implications for planetary accretion and differentiation}}.
\newblock \bibinfo{journal}{Chemie der Erde / Geochemistry}
  \bibinfo{volume}{69}, \bibinfo{pages}{101--125}.
\newblock \DOIprefix\doi{10.1016/j.chemer.2008.10.001}.
%Type = Article
\bibitem[{{Walsh} et~al.(2011){Walsh}, {Morbidelli}, {Raymond}, {O'Brien} and
  {Mandell}}]{Walsh2011}
\bibinfo{author}{{Walsh}, K.J.}, \bibinfo{author}{{Morbidelli}, A.},
  \bibinfo{author}{{Raymond}, S.N.}, \bibinfo{author}{{O'Brien}, D.P.},
  \bibinfo{author}{{Mandell}, A.M.}, \bibinfo{year}{2011}.
\newblock \bibinfo{title}{{A low mass for Mars from Jupiter's early gas-driven
  migration}}.
\newblock \bibinfo{journal}{Nature} \bibinfo{volume}{475},
  \bibinfo{pages}{206--209}.
\newblock \DOIprefix\doi{10.1038/nature10201},
  \href{http://arxiv.org/abs/1201.5177}{\tt arXiv:1201.5177}.
%Type = Article
\bibitem[{Warren(2011)}]{WARREN201193}
\bibinfo{author}{Warren, P.H.}, \bibinfo{year}{2011}.
\newblock \bibinfo{title}{Stable-isotopic anomalies and the accretionary
  assemblage of the earth and mars: A subordinate role for carbonaceous
  chondrites}.
\newblock \bibinfo{journal}{Earth and Planetary Science Letters}
  \bibinfo{volume}{311}, \bibinfo{pages}{93--100}.
\newblock \URLprefix
  \url{https://www.sciencedirect.com/science/article/pii/S0012821X11005115},
  \DOIprefix\doi{https://doi.org/10.1016/j.epsl.2011.08.047}.
%Type = Article
\bibitem[{{Weidenschilling} et~al.(1997){Weidenschilling}, {Spaute}, {Davis},
  {Marzari} and {Ohtsuki}}]{weidenschilling97}
\bibinfo{author}{{Weidenschilling}, S.J.}, \bibinfo{author}{{Spaute}, D.},
  \bibinfo{author}{{Davis}, D.R.}, \bibinfo{author}{{Marzari}, F.},
  \bibinfo{author}{{Ohtsuki}, K.}, \bibinfo{year}{1997}.
\newblock \bibinfo{title}{{Accretional Evolution of a Planetesimal Swarm}}.
\newblock \bibinfo{journal}{Icarus} \bibinfo{volume}{128},
  \bibinfo{pages}{429--455}.
\newblock \DOIprefix\doi{10.1006/icar.1997.5747}.
%Type = Article
\bibitem[{{Wetherill}(1990)}]{Wetherill1990}
\bibinfo{author}{{Wetherill}, G.W.}, \bibinfo{year}{1990}.
\newblock \bibinfo{title}{{Formation of the earth}}.
\newblock \bibinfo{journal}{Annual Review of Earth and Planetary Sciences}
  \bibinfo{volume}{18}, \bibinfo{pages}{205--256}.
\newblock \DOIprefix\doi{10.1146/annurev.ea.18.050190.001225}.
%Type = Article
\bibitem[{{Wetherill} and {Stewart}(1993)}]{wetherill93}
\bibinfo{author}{{Wetherill}, G.W.}, \bibinfo{author}{{Stewart}, G.R.},
  \bibinfo{year}{1993}.
\newblock \bibinfo{title}{{Formation of Planetary Embryos: Effects of
  Fragmentation, Low Relative Velocity, and Independent Variation of
  Eccentricity and Inclination}}.
\newblock \bibinfo{journal}{Icarus} \bibinfo{volume}{106},
  \bibinfo{pages}{190--209}.
\newblock \DOIprefix\doi{10.1006/icar.1993.1166}.
%Type = Article
\bibitem[{{Wisdom} and {Holman}(1992)}]{Wisdom92}
\bibinfo{author}{{Wisdom}, J.}, \bibinfo{author}{{Holman}, M.},
  \bibinfo{year}{1992}.
\newblock \bibinfo{title}{{Symplectic Maps for the n-Body Problem: Stability
  Analysis}}.
\newblock \bibinfo{journal}{Astronomical Journal} \bibinfo{volume}{104},
  \bibinfo{pages}{2022}.
\newblock \DOIprefix\doi{10.1086/116378}.
%Type = Article
\bibitem[{Woo et~al.(2023)Woo, Morbidelli, Grimm, Stadel and
  Brasser}]{Woo_2023}
\bibinfo{author}{Woo, J.}, \bibinfo{author}{Morbidelli, A.},
  \bibinfo{author}{Grimm, S.}, \bibinfo{author}{Stadel, J.},
  \bibinfo{author}{Brasser, R.}, \bibinfo{year}{2023}.
\newblock \bibinfo{title}{Terrestrial planet formation from a ring}.
\newblock \bibinfo{journal}{Icarus} \bibinfo{volume}{396},
  \bibinfo{pages}{115497}.
\newblock \URLprefix \url{http://dx.doi.org/10.1016/j.icarus.2023.115497},
  \DOIprefix\doi{10.1016/j.icarus.2023.115497}.
%Type = Article
\bibitem[{{Worsham} and {Kleine}(2021)}]{Worsham2021}
\bibinfo{author}{{Worsham}, E.A.}, \bibinfo{author}{{Kleine}, T.},
  \bibinfo{year}{2021}.
\newblock \bibinfo{title}{{Late accretionary history of Earth and Moon
  preserved in lunar impactites}}.
\newblock \bibinfo{journal}{Science Advances} \bibinfo{volume}{7},
  \bibinfo{pages}{eabh2837}.
\newblock \DOIprefix\doi{10.1126/sciadv.abh2837}.
%Type = Article
\bibitem[{{Zhu} et~al.(2021){Zhu}, {Morbidelli}, {Neumann}, {Yin}, {Day},
  {Rubie}, {Archer}, {Artemieva}, {Becker} and {W{\"u}nnemann}}]{zhu21}
\bibinfo{author}{{Zhu}, M.H.}, \bibinfo{author}{{Morbidelli}, A.},
  \bibinfo{author}{{Neumann}, W.}, \bibinfo{author}{{Yin}, Q.Z.},
  \bibinfo{author}{{Day}, J.M.D.}, \bibinfo{author}{{Rubie}, D.C.},
  \bibinfo{author}{{Archer}, G.J.}, \bibinfo{author}{{Artemieva}, N.},
  \bibinfo{author}{{Becker}, H.}, \bibinfo{author}{{W{\"u}nnemann}, K.},
  \bibinfo{year}{2021}.
\newblock \bibinfo{title}{{Common feedstocks of late accretion for the
  terrestrial planets}}.
\newblock \bibinfo{journal}{Nature Astronomy} \bibinfo{volume}{5},
  \bibinfo{pages}{1286--1296}.
\newblock \DOIprefix\doi{10.1038/s41550-021-01475-0}.

\end{thebibliography}

\end{document}